\documentclass[aps,prd,superscriptaddress,notitlepage,showkeys,showpacs,10pt]{revtex4-1}

\usepackage[english]{babel}
\usepackage{amsmath,bm}
\usepackage{amssymb}
\usepackage{enumitem}
\usepackage{textcomp}
\usepackage{graphicx}
\usepackage{dsfont}
\usepackage{color}
\usepackage{hyperref}
\definecolor{purple}{rgb}{0.8,0,0.6}

\begin{document}
	
\title{Influence of backreaction of electric fields and Schwinger effect on inflationary magnetogenesis}

\author{O.O.~Sobol}
\affiliation{Institute of Physics, Laboratory for Particle Physics and Cosmology, \'{E}cole Polytechnique F\'{e}d\'{e}rale de Lausanne, CH-1015 Lausanne, Switzerland}
\affiliation{Physics Faculty, Taras Shevchenko National University of Kyiv, 64/13, Volodymyrska Street, 01601 Kyiv, Ukraine}

\author{E.V.~Gorbar}
\affiliation{Physics Faculty, Taras Shevchenko National University of Kyiv, 64/13, Volodymyrska Street, 01601 Kyiv, Ukraine}
\affiliation{Bogolyubov Institute for Theoretical Physics, 14-b, Metrologichna Street, 03143 Kyiv, Ukraine}

\author{M.~Kamarpour}
\affiliation{Physics Faculty, Taras Shevchenko National University of Kyiv, 64/13, Volodymyrska Street, 01601 Kyiv, Ukraine}

\author{S.I.~Vilchinskii}
\affiliation{Physics Faculty, Taras Shevchenko National University of Kyiv, 64/13, Volodymyrska Street, 01601 Kyiv, Ukraine}
\affiliation{D\'{e}partement de Physique Th\'{e}orique, Center for Astroparticle Physics, Universit\'{e} de Gen\`{e}ve, 1211 Gen\`{e}ve 4, Switzerland}

\date{\today}

\begin{abstract}
We study the generation of electromagnetic fields during inflation when the conformal invariance of Maxwell's action is broken by the kinetic coupling $f^{2}(\phi)F_{\mu\nu}F^{\mu\nu}$ of the electromagnetic field to the inflaton field $\phi$. We consider the case where the coupling function $f(\phi)$ decreases in time during inflation and, as a result, the electric component of the energy density dominates over the magnetic one. The system of equations which governs the joint evolution of the scale factor, inflaton field, and electric energy density is derived. 
The backreaction occurs when the electric energy density becomes as large as the product of the slow-roll parameter $\epsilon$ and inflaton energy density, $\rho_{E}\sim \epsilon \rho_{\rm inf}$. It affects the inflaton field evolution and leads to the scale-invariant electric power spectrum and the magnetic one which is blue with the spectral index $n_{B}=2$ for any decreasing coupling function. This gives an upper limit on the present-day value of observed magnetic fields below $10^{-22}\,{\rm G}$. It is worth emphasizing that since the effective electric charge of particles $e_{\rm eff}=e/f$ is suppressed by the coupling function, the Schwinger effect becomes important only at the late stages of inflation when the inflaton field is close to the minimum of its potential. The Schwinger effect abruptly decreases the value of the electric field, helping to finish the inflation stage and enter the stage of preheating. It effectively produces the charged particles, implementing the Schwinger reheating scenario even before the fast oscillations of the inflaton. The numerical analysis is carried out in the Starobinsky model of inflation for the powerlike $f\propto a^{\alpha}$ and Ratra-type $f=\exp(\beta\phi/M_{p})$ coupling functions.
\end{abstract}


\maketitle

\section{Introduction}
\label{sec-intro}

One of the important problems of modern cosmology is the origin of the magnetic fields which are present at all scales in the Universe \cite{Kronberg:1994,Grasso:2001,Widrow:2002,Giovannini:2004,Kandus:2011,Durrer:2013,Subramanian:2016,Giovannini:2018b}, especially of the magnetic fields detected in the cosmic voids through the gamma-ray observations of distant blazars \cite{Neronov:2010,Tavecchio:2010,Taylor:2011,Dermer:2011,Caprini:2015} with very large coherence scale $\lambda_{B}\gtrsim 1\,$Mpc. Together with the observations of the cosmic microwave background (CMB) \cite{Planck:2015-pmf,Sutton:2017,Jedamzik:2018,Paoletti:2018,Giovannini:2018b} and ultrahigh-energy cosmic rays \cite{Bray:2018} this implies the following constraints on the strength of these fields: $10^{-18}\lesssim B_{0}\lesssim 10^{-9}\,$G. If the correlation length of the magnetic field is $\lambda_{B}\lesssim 1\,$Mpc, the minimal needed magnetic field strength is larger by the factor $(\lambda_{B}/1\,{\rm Mpc})^{-1/2}$ \cite{Neronov:2010,Taylor:2011,Dermer:2011,Caprini:2015}.

Observed intergalactic magnetic fields can have either astrophysical or primordial origin, and both magnetogenesis scenarios are currently under discussion. Although astrophysical mechanisms based on a Biermann battery \cite{Biermann:1950,Pudritz:1989,Gnedin:2000} have been proposed to generate the ``seed'' magnetic fields and different types of dynamo can enhance them \cite{Zeldovich:1980book,Lesch:1995,Kulsrud:1997,Colgate:2001}, it is problematic to embed the magnetic fields with a large correlation length into the cosmic voids. Therefore, the primordial origin of the large scale magnetic fields seems to be more realistic. In particular, the phase transitions in the early Universe may lead to the magnetic fields of the necessary strength \cite{Hogan:1983,Quashnock:1989,Vachaspati:1991,Cheng:1994,Sigel:1997,Ahonen:1998}. However, their coherence length is determined by the horizon size at the moment of phase transition and is much less than Mpc today. Then the most natural is the inflationary magnetogenesis, proposed in Refs.~\cite{Turner:1988,Ratra:1992}, which can easily attain
very large coherence length. 

Since Maxwell's action is conformally invariant, the fluctuations of the electromagnetic field do not undergo enhancement in the conformally flat inflationary background \cite{Parker:1968}. In order to generate electromagnetic fields we need to break the conformal invariance. This can be done by introducing the interaction with scalar or pseudoscalar inflaton fields or curvature scalar (see, e.g., the pioneer works \cite{Turner:1988,Ratra:1992,Garretson:1992,Dolgov:1993}). In our study we consider the kinetic coupling of the electromagnetic field to the scalar inflaton via the term $f^{2}(\phi)F_{\mu\nu}F^{\mu\nu}$, which was first introduced by Ratra \cite{Ratra:1992} and has been revisited many times for different types of coupling functions \cite{Giovannini:2001,Bamba:2004,Martin:2008,Demozzi:2009,Kanno:2009,Ferreira:2013,Ferreira:2014,Vilchinskii:2017,Sharma:2017b}.

This model modifies the standard electromagnetic Lagrangian, multiplying it by the function of the inflaton field. As it was mentioned in \cite{Demozzi:2009}, if one rescales the electromagnetic potential in order to absorb $f^{2}$, electric charges of particles effectively will depend on $f^{-1}$. Therefore, to avoid the strong coupling problem during inflation one needs to require $f\ge 1$. Since the inflaton field and the scale factor change monotonously during inflation it is natural to assume that the coupling function is a decreasing function during inflation which attains large values in the beginning.

For decreasing coupling functions, it is well known that the electric energy density dominates the magnetic one \cite{Martin:2008,Demozzi:2009,Vilchinskii:2017} and if we try to generate the magnetic field strong enough to be in accord with the observations, the electric field appears to be even stronger and its energy density exceeds that of the inflaton field. This is known as the backreaction problem. In previous studies, the authors tried to avoid this problem, because it does not allow us to solve the background equations and Maxwell equation separately \cite{Martin:2008,Demozzi:2009,Ferreira:2013,Ferreira:2014}. However, it is interesting what happens when the backreaction becomes important and whether the amplification really stops in this regime. These are open questions in the literature. 

Since strong electric fields could be generated during inflation, the pair creation in a strong electric field, which is known as the 
Schwinger effect \cite{Schwinger:1951}, can become important and affect the magnetogenesis. The
Schwinger effect in the constant and homogeneous background electric field in de Sitter space-time was investigated by many authors~\cite{Froeb:2014,Kobayashi:2014,Bavarsad:2016,Stahl:2016a,Stahl:2016b,Hayashinaka:2016a,Hayashinaka:2016b,Sharma:2017,Tangarife:2017,Bavarsad:2018,Hayashinaka:2018,Hayashinaka:thesis,Stahl:2018, Geng:2018}. The cases of (1+1)-dimensional \cite{Froeb:2014,Stahl:2016a,Stahl:2016b}, (2+1)-dimensional \cite{Bavarsad:2016}, and (3+1)-dimensional \cite{Kobayashi:2014,Hayashinaka:2016a,Hayashinaka:2016b,Hayashinaka:2018,Hayashinaka:thesis} de Sitter space-time with scalar \cite{Froeb:2014,Kobayashi:2014,Bavarsad:2016,Hayashinaka:2016b,Hayashinaka:2018,Hayashinaka:thesis,Stahl:2018} and spinor charged fields \cite{Stahl:2016a,Stahl:2016b,Hayashinaka:2016a,Hayashinaka:2018,Hayashinaka:thesis}, including also an external magnetic field \cite{Bavarsad:2018}, were investigated. It is important to note that constant electric energy density is considered rather than the case of a constant comoving electric field. According to Ref.~\cite{Giovannini:2018a}, maintaining this regime would require the existence of \textit{ad hoc} currents that could violate the second law of thermodynamics.

The cosmological Schwinger effect relates the particle production by an electric field and the exponential expansion of the Universe and contains interesting features which are absent in its flat-space counterpart, namely (i) the infrared hyperconductivity in the bosonic case when the conductivity becomes very large in the limit of small mass of charged particles and (ii) the negative conductivity in the weak-field regime $eE\ll H^{2}$ which can, in principle, lead to the enhancement of the electric field.

The induced current of created particles obtained by direct averaging of the corresponding current operator contains ultraviolet divergences, which can be regularized using adiabatic subtraction \cite{Kobayashi:2014,Hayashinaka:2016a} or the point-splitting method \cite{Hayashinaka:2016b}. Although these techniques remove the divergent parts, the finite part is not uniquely defined and depends on the applied subtraction scheme. Adiabatic subtraction to the second order in the adiabaticity parameter and point-splitting regularization define the minimal subtraction scheme which is commonly used. However, in the massive limit, when the particle's mass is much greater than the Hubble parameter, the current and conductivity contain exponentially unsuppressed terms, which is not consistent with the standard Bogolyubov calculations. In order to eliminate this discrepancy the authors of Ref.~\cite{Hayashinaka:2018} proposed a new maximal subtraction scheme, which normalizes the behavior of the current in the massive case. However, the main features in the small mass regime remain the same and the infrared hyperconductivity in the scalar case is observed \cite{Kobayashi:2014,Hayashinaka:2016b}. 

In addition, the charged particles show negative conductivity in the weak-field regime, which can lead to an instability and an avalanchelike enhancement of electric field up to a certain critical value \cite{Hayashinaka:2016a,Hayashinaka:thesis}. In the very recent article \cite{Stahl:2018} Stahl considers a possibility to enhance the quantum fluctuations of the electromagnetic field even without interaction with an inflaton. However, the negative conductivity is rather speculative \cite{Hayashinaka:2018,Hayashinaka:thesis} and may be an artifact of used subtraction schemes.

An attempt to combine the generation of the electromagnetic field due to kinetic coupling with the inflaton, the Schwinger effect and the backreaction into a consistent picture was made in Ref.~\cite{Kitamoto:2018}. It was shown that the expressions for the Schwinger current in the time-dependent electric background in the strong-field regime have the same functional dependence as in the case of a constant electric field. However, the backreaction on the background evolution was calculated perturbatively only in the first order, which is valid only at the early stages of inflation. The impact on the generated magnetic field was not discussed as well.

This paper is organized as follows. We derive a self-consistent system of equations which describes the joint evolution of the scale factor, inflaton field, and electric field energy density in Sec.~\ref{sec-basics}, where we take into account the backreaction of generated fields on the background evolution and the Schwinger effect which is important at the late stages of inflation. In Sec.~\ref{sec-BR-spectra}, we study the regime when the backreaction becomes important and analyze the main features of the electromagnetic field power spectra generated in this regime. The results of numerical calculations of the power spectra of generated fields and the present value of the magnetic field in the Starobinsky inflation model for two types of coupling functions are represented in Sec.~\ref{sec-numerical}. The summary of the obtained results is given in Sec.~\ref{sec-concl}.

\section{Self-consistent system of equations}
\label{sec-basics}

We consider a spatially flat Friedmann-Lama\^{i}tre-Robertson-Walker Universe with metric tensor
\begin{equation}
\label{metric}
g_{\mu\nu}={\rm diag}\,(1,\,-a^{2},\,-a^{2},\,-a^{2}), \quad \sqrt{-g}=a^{3},
\end{equation}
and use the natural system of units where $\hbar=c=1$, $M_{p}=(8\pi G)^{-1/2}=2.4\cdot 10^{18}\,{\rm GeV}$ is a reduced Planck mass, and $e=\sqrt{4\pi\alpha}\approx 0.3$ is the absolute value of the electron's charge.

The action which describes inflaton field $\phi$, electromagnetic field $A_{\mu}$, and charged field $\chi$ (either bosonic or fermionic) reads
\begin{equation}
\label{action}
S=\int d^{4}x \sqrt{-g}\left[\frac{1}{2}g^{\mu\nu}\partial_{\mu}\phi\partial_{\nu}\phi-V(\phi)-\frac{1}{4}f^{2}(\phi)g^{\mu\alpha}g^{\nu\beta}F_{\mu\nu}F_{\alpha\beta}+\mathcal{L}_{\rm charged}(A,\chi)\right],
\end{equation}
where $V(\phi)$ is the inflaton effective potential; $f(\phi)$ is the kinetic coupling function; and
$\mathcal{L}_{\rm charged}(A,\chi)$ is a gauge-invariant Lagrangian of the charged field $\chi$ interacting with the electromagnetic field $A_{\mu}$.

The corresponding Euler-Lagrange equations have the form
\begin{equation}
\label{KGF}
\frac{1}{\sqrt{-g}}\partial_{\mu}\left[\sqrt{-g}g^{\mu\nu}\partial_{\nu}\phi\right]+\frac{dV}{d\phi}=-\frac{1}{2}ff'F_{\mu\nu}F^{\mu\nu},
\end{equation}
\begin{equation}
\label{Maxwell}
\frac{1}{\sqrt{-g}}\partial_{\mu}\left[\sqrt{-g}g^{\mu\alpha}g^{\nu\beta}f^{2}(\phi)F_{\alpha\beta}\right]=-j^{\nu},
\end{equation}
where the 4-current is defined as usual
\begin{equation}
\label{4-current}
j^{\mu}=\frac{\partial\mathcal{L}_{\rm charged}(A,\chi)}{\partial A_{\mu}}.
\end{equation}
The right-hand side of Eq.~(\ref{KGF}) describes the backreaction of created electric fields on the evolution of the inflaton 
field.

Evolution of the Universe is driven by the total energy density of all fields. It can be calculated as the $00$-component of the 
stress-energy tensor. The latter is defined as usual as
\begin{equation}
T_{\mu\nu}=\frac{2}{\sqrt{-g}}\frac{\delta S}{\delta g^{\mu\nu}}=\partial_{\mu}\phi\partial_{\nu}\phi-f^{2}(\phi)g^{\alpha\beta}F_{\mu\alpha}F_{\nu\beta}-g_{\mu\nu} \mathcal{L}_{0}+T_{\mu\nu}^{({\rm charged})},
\end{equation}
where $\mathcal{L}_{0}$ is the Lagrangian density of the inflaton and electromagnetic fields which is represented by the first three 
terms in the square brackets in Eq.~(\ref{action}).

In the simplest case the inflaton field is spatially homogeneous. If the coupling function $f(\phi)$ always decreases in time, then it is well known \cite{Martin:2008,Demozzi:2009} that the electric component of the created electromagnetic field dominates the magnetic one and leads to the backreaction problem. Therefore, we take into account the presence of electric field $F_{0i}=a(t)E_{i}(t)$ and neglect the magnetic component $F_{ij}=a^{2}\varepsilon_{ijk}B_{k}\approx 0$. Then the energy density reads
\begin{equation}
\label{energy-density}
\rho=\left[\frac{1}{2}\dot{\phi}^{2}+V(\phi)\right]+\frac{1}{2}f^{2}E^{2}+\rho_{\chi}=\rho_{\rm inf}+\rho_{E}+\rho_{\chi},
\end{equation}
where $\rho_{\chi}$ is the energy density of the charged particles produced by the Schwinger effect.

The Friedmann, Klein-Gordon, and Maxwell equations take the following form:
\begin{eqnarray}
\label{Friedmann}
&&H^{2}=\left(\frac{\dot{a}}{a}\right)^{2}=\frac{1}{3M_{p}^{2}}\left(\rho_{\rm inf}+\rho_{E}+\rho_{\chi}\right),\\
\label{KGF-2}
&&\ddot{\phi}+3H\dot{\phi}+\frac{dV}{d\phi}=f(\phi)f'(\phi)E^{2}(t),\\
\label{Maxwell-2}
&&\partial_{t}\left(a^{2}f^{2}E_{i}\right)=-aj_{i},
\end{eqnarray}
where overdots denote the derivatives with respect to cosmic time and the prime denotes the derivative with respect to the inflaton field. 

It has to be mentioned that we consider the classical electric field, which is generated from quantum fluctuations due to interaction with the inflaton. It is useful to analyze the mode composition of this field. When the modes are inside the horizon they oscillate in time and have to be treated as quantum fluctuations. However, when they cross the horizon the mode functions start to behave monotonically and can be chosen real. According to Ref.~\cite{Lyth:2008}, this corresponds to the quantum-to-classical transition and these modes can be treated classically. Therefore, the contribution to the electric field is made only by the modes, which are outside the horizon. Since their wavelength is larger than the horizon, i.e. the largest observable scale, the corresponding electric field can be considered homogeneous.

\subsection{Equation for electric field energy density}

It is convenient to rewrite the Maxwell equation (\ref{Maxwell-2}) in terms of electric field energy density $\rho_{E}=f^{2}E^{2}/2$:
\begin{equation}
\label{Maxwell-3}
\dot{\rho}_{E}+4H\rho_{E}+2\frac{\dot{f}}{f}\rho_{E}=-\frac{1}{a}\left(E\cdot j\right).
\end{equation}

The ``classical'' part of the electric energy density at a certain moment of time is determined by the modes, which crossed the horizon from the beginning of inflation till the moment under consideration:
\begin{equation}
\rho_{E}(t)=\int_{k_{i}}^{k_{H}(t)}dk\frac{d\rho_{E}}{dk}(t), \quad k_{H}(t)=a(t)H(t),
\end{equation}
where $k_{i}$ is the momentum of the mode which crosses the horizon at the beginning of inflation.

However, Eq.~(\ref{Maxwell-3}) does not take into account the fact that the number of relevant modes with wavelength larger than the horizon changes in time. In order to deal with this, we should introduce an additional term which describes the modes crossing the horizon at a given time $t$ and starting to contribute to the total energy density of the electric field:
\begin{equation}
\left(\dot{\rho}_{E}\right)_{H}=\left.\frac{d\rho_{E}}{dk}\right|_{k=k_{H}}\cdot \frac{dk_{H}}{dt}.
\label{boundary-term}
\end{equation} 

The power spectra of electric and magnetic fields are expressed through the mode function of the electromagnetic field
$\mathcal{A}(k,t)$ in the standard way \cite{Martin:2008}:
\begin{eqnarray}
\label{E-power-spectrum}
\frac{d\rho_{E}}{dk}&=&\frac{f^{2}}{2\pi^{2}}\frac{k^{2}}{a^{2}}\left|\frac{\partial}{\partial t}\frac{\mathcal{A}(k,t)}{f(t)}\right|^{2},\\
\label{B-power-spectrum}
\frac{d\rho_{B}}{dk}&=&\frac{1}{2\pi^{2}}\frac{k^{4}}{a^{4}}\left|\mathcal{A}(k,t)\right|^{2}.
\end{eqnarray}

This mode function satisfies the following equation (in the conformal time $\eta$):
\begin{equation}
\label{mode-eq}
\frac{\partial^{2}\mathcal{A}_{k}}{\partial \eta^{2}}+\left(k^{2}-\frac{1}{f}\frac{\partial^{2} f}{\partial \eta^{2}}\right)\mathcal{A}_{k}=0,
\end{equation}
and the initial conditions for the modes inside the horizon have the form of the Bunch-Davies vacuum
\begin{equation}
\label{init-BD}
\mathcal{A}(k,t)=\frac{1}{\sqrt{2k}}e^{-i k\eta(t)}, \quad k\eta(t)\to -\infty.
\end{equation}

When a certain mode crosses the horizon it changes its dependence on time from oscillatory to monotonous. We can assume that at the 
moment of horizon crossing its behavior is still approximately described by Eq.~(\ref{init-BD}). Then
\begin{equation}
\left|\frac{\partial}{\partial t}\frac{\mathcal{A}(k,t)}{f(t)}\right|^{2}\approx \frac{1}{2k f^{2}}\left[\frac{k^{2}}{a^{2}}+\left(\frac{\dot{f}}{f}\right)^{2}\right]
\end{equation}
and the ``boundary'' term (\ref{boundary-term}) takes the form
\begin{equation}
\label{boundary}
\left(\dot{\rho}_{E}\right)_{H}=\left.\frac{f^{2}}{2\pi^{2}}\frac{k^{2}}{a^{2}}\frac{1}{2k f^{2}}\left[\frac{k^{2}}{a^{2}}+\left(\frac{\dot{f}}{f}\right)^{2}\right]\right|_{k=k_{H}}\cdot \frac{dk_{H}}{dt}= \frac{H^{3}}{4\pi^{2}}\left[H^{2}+\left(\frac{\dot{f}}{f}\right)^{2}\right]
\end{equation}

Finally, the equation which governs the behavior of electric field energy density is given by
\begin{equation}
\label{Maxwell-4}
\dot{\rho}_{E}+4H\rho_{E}+2\frac{\dot{f}}{f}\rho_{E}=-\frac{1}{a}\left(E\cdot j\right)+\frac{H^{3}}{4\pi^{2}}\left[H^{2}+\left(\frac{\dot{f}}{f}\right)^{2}\right].
\end{equation}

It is possible to derive Eq.~(\ref{Maxwell-4}) in an alternative way by inspecting the time dependence of the electric power spectrum. In the deeply subhorizon regime $k\gg aH$ the mode function takes the value of the Bunch-Davies initial conditions (\ref{init-BD}). Far outside the horizon $k\ll aH$, the solution of Eq.~(\ref{mode-eq}) is
\begin{equation}
\label{outside-horizon}
\frac{\mathcal{A}_{k}}{f}=C_{1}+C_{2}\int_{t_{k}}^{t}\frac{dt'}{a(t')f^{2}(t')}, \quad\frac{\partial}{\partial t}\frac{\mathcal{A}_{k}}{f}=\frac{C_{2}}{af^{2}}.
\end{equation}

Matching solutions (\ref{init-BD}) and (\ref{outside-horizon}) at the moment of horizon crossing $t_{k}$ when $k=aH$, we find
\begin{equation}
C_{1}=\frac{e^{-ik\eta_{k}}}{\sqrt{2k}f_{k}},\quad C_{2}=-\frac{e^{-ik\eta_{k}}a_{k}f_{k}}{\sqrt{2k}}\left[iH_{k}+\frac{\dot{f}_{k}}{f_{k}}\right],
\end{equation}
where all the quantities with subscript $k$ must be taken at the moment of horizon crossing $t_{k}$. Then, using
Eqs.~(\ref{E-power-spectrum})--(\ref{B-power-spectrum}), we can write the power spectra of generated fields as follows:
\begin{eqnarray}
\label{E-power-spectrum-2}
\frac{d\rho_{E}}{d\ln\,k}&=&\frac{1}{2\pi^{2}a^{4}f^{2}}k^{3}|C_{2}(k)|^{2}=\frac{k^{4}}{4\pi^{2}a^{4}}\frac{f_{k}^{2}}{f^{2}}\left[1+\left(\frac{1}{H_{k}}\frac{\dot{f}_{k}}{f_{k}}\right)^{2}\right],\\
\label{B-power-spectrum-2}
\frac{d\rho_{B}}{d\ln\,k}&=&\frac{1}{2\pi^{2}a^{4}}f^{2}k^{5}\left|C_{1}(k)+C_{2}(k)\int_{t_{k}}^{t}\frac{dt'}{a(t')f^{2}(t')}\right|^{2}=\nonumber\\
&=&\frac{k^{4}}{4\pi^{2}a^{4}}\frac{f^{2}}{f_{k}^{2}}\left|1-\left(i+\frac{1}{H_{k}}\frac{\dot{f}_{k}}{f_{k}}\right)H_{k}\int_{t_{k}}^{t}\frac{a_{k}f_{k}^{2}}{a(t')f^{2}(t')}dt'\right|^{2}.
\end{eqnarray}

Integrating Eq.~(\ref{E-power-spectrum-2}) over modes outside the horizon at a given moment of time, we obtain the energy density of the
electric field:
\begin{equation}
\rho_{E}(t)=\frac{1}{a^{4}f^{2}}\int_{k_{i}}^{k_{H}(t)}S(k),\quad S(k)=\frac{k^{3}f_{k}^{2}}{4\pi^{2}}\left[1+\left(\frac{1}{H_{k}}\frac{\dot{f}_{k}}{f_{k}}\right)^{2}\right],
\end{equation}
where $S(k)$ is a function which only depends on momentum $k$ and is time independent. Then, obviously, the time derivative of the energy density equals
\begin{equation}
\dot{\rho}_{E}=-4H \rho_{E}-2\frac{\dot{f}}{f}\rho_{E}+\frac{1}{a^{4}f^{2}}S(k_{H})\frac{dk_{H}}{dt}.
\end{equation}
Taking into account that $t_{k}(k=k_{H})\equiv t$, we obtain Eq.~(\ref{Maxwell-4}) without the Schwinger term, which was also absent in Eq.~(\ref{mode-eq}) and could be added in the final equation phenomenologically.

However, it is impossible to derive a similar equation for the magnetic energy density, because the terms proportional to $C_{1}$ and $C_{2}$ in Eq.~(\ref{B-power-spectrum-2}), in general, have different behaviors in time and their relative contribution could change during the time evolution. In other words, the time dependence of the magnetic power spectrum cannot be extracted in the form of a universal function like in the case of electric field energy density (\ref{E-power-spectrum-2}).

At the moment of horizon crossing there is only the term in Eq.~(\ref{B-power-spectrum-2}), proportional to $C_{1}$ because the integral vanishes.
If $af^2$ is an increasing function, then the second term is a decaying mode and it can be neglected. In this case it is possible to obtain the equation for the magnetic energy density. However it is not interesting, because the coupling function decreases slower than $a^{-1/2}$ and the generated electromagnetic fields are too small to explain the observational data \cite{Martin:2008}. We do not take into account the possibility of a growing coupling function because it causes the strong coupling problem in the beginning of inflation.

To derive the boundary term in Eq.~(\ref{Maxwell-4}) we used the decomposition of the electromagnetic field operator over the set of Fourier modes. However, in the presence of the Schwinger current, strictly speaking, this is not advantageous because this current has a nonlinear dependence on the electric field. Nevertheless, we will show in the following sections that the Schwinger effect is important only at the late stages of inflation when the electric field energy density is large and the boundary term is negligible in comparison with other terms in Eq.~(\ref{Maxwell-4}). On the contrary, at the beginning of inflation when the electric energy density is close to zero, the boundary term is very important because it generates the initial value of the electric field which will be enhanced due to the interaction with the inflaton. At that time the Schwinger effect is negligibly small and can be excluded from the consideration. This justifies our derivation.

\subsection{Schwinger effect}
\label{subsec-Schwinger}

To close the system of equations we need two other equations which determine the Schwinger current $j$ and the energy density of particles $\rho_{\chi}$ created via the Schwinger process.

We will use the expressions for the Schwinger currents derived in the minimal subtraction scheme in Refs.~\cite{Kobayashi:2014,Hayashinaka:2016a}. The general expressions are rather cumbersome. Therefore, it is more convenient to use their asymptotics governed by the 
parameters
\begin{equation}
M=\frac{m}{H}, \quad L=\frac{eE}{H^{2}}=\frac{e}{f}\sqrt{\frac{2\rho_{E}}{H^{4}}},
\end{equation}
where $m$ is a charged particle's mass, and the effective charge $e_{\rm eff}=e/f$ naturally appears in the second expression.

Since the electric energy density always satisfies $\rho_{E}\lesssim \rho_{\rm tot}=3H^{2}M_{p}^{2}$ when the backreaction is taken into account, the parameter $L$ is damped by the effective charge during almost the entire period of inflation and can become large only at the end, when the inflaton field quickly rolls down to the minimum of its potential and $f\to 1$. The typical value of the Hubble parameter, which can be fixed from the observations of the amplitude of primordial perturbations \cite{Planck:2018-infl}, is $H\sim 10^{-5}\,M_{p}\sim 10^{13}\,{\rm GeV}$. On the other hand, the lightest charged spin-$1/2$ particle, the electron, has a mass $m\sim 10^{-3}\,{\rm GeV}$ and the lightest charged scalar particle, the pion, has a mass $m\sim 0.1\,{\rm GeV}$. Therefore, we are interested in the small mass limit $M\ll 1$.

In the weak-field regime, $L\ll 1$, the fermionic Schwinger current has the form \cite{Hayashinaka:2016a}
\begin{equation}
\label{current-fermi}
j_{f}= \frac{aH^{3}}{18\pi^{2}}\left(6 \ln \frac{m}{H}+6\gamma_{E}-1\right)\frac{e^{2}E}{H^{2}},
\end{equation}
where $\gamma_{E}\approx -0.577$ is the Euler's constant. It is important to note that the current is always negative in the low mass limit $m<H$ and, in principle, can lead to the enhancement of the electric field \cite{Stahl:2018}.

In the bosonic case the current is positive; however, it has a very large value in the low mass regime as it diverges like $1/M^{2}$ at $M\to 0$ \cite{Kobayashi:2014,Hayashinaka:2016b}:
\begin{equation}
j_{b}= \frac{3 aH^{3}}{4\pi^{2}}\frac{e^{2}E}{m^{2}}.
\end{equation}
In the literature this phenomenon is often called the infrared hyperconductivity. It has to be mentioned that this expression is valid in the extremely low field regime, when $L\ll M\ll 1$. When $L$ becomes larger than $M$, the current quickly diminishes to the values comparable with the fermionic case (\ref{current-fermi}) and in the region $L\sim 1$ the negative conductivity can be observed \cite{Hayashinaka:2016b}.

Expressing the dissipation term on the right-hand side of Eq.~(\ref{Maxwell-4}) in terms of $\rho_{E}$, we obtain
\begin{eqnarray}
\frac{1}{a}(j_{b}\cdot E)&=& \frac{3}{2\pi^{2}}\frac{e^{2}}{f^{2}}\frac{H^{3}}{m^{2}}\rho_{E}, \label{weak-field-bose}\\
\frac{1}{a}(j_{f}\cdot E)&=& \frac{1}{9\pi^{2}}\frac{e^{2}}{f^{2}} \left(6 \ln \frac{m}{H}+6\gamma_{E}-1\right) H\rho_{E}.\label{weak-field-fermi}
\end{eqnarray}

It has to be compared with the main terms in Eq.~(\ref{Maxwell-4}) which are of order $H\rho_{E}$. For scalar particles, the weak-field expression (\ref{weak-field-bose}) becomes important when
\begin{equation}
\label{ineq-1}
\frac{e}{\pi f}\gtrsim  \frac{m}{H}.
\end{equation}
The corresponding $L$-parameter must be smaller than $M$, which gives
\begin{equation}
\label{ineq-2}
L=\frac{e}{f}\frac{\sqrt{2\rho_{E}}}{H^{2}}\lesssim \frac{m}{H}.
\end{equation}
This implies that at the moment when expression (\ref{weak-field-bose}) may be important the electric energy density $\rho_{E}\sim H^{4}\ll \rho_{\rm tot}$. This can take place only at the early stages of inflation when the coupling function is very large. Therefore, the two inequalities (\ref{ineq-1}) and (\ref{ineq-2}) can be satisfied simultaneously either for the extremely low masses or in the case where the coupling function changes not very strongly during inflation. Both cases are unfavorable from the point of view of magnetogenesis as the generated magnetic fields would be very weak: in the first case the magnetogenesis terminates too early, and in the second case the enhancement can be insufficiently strong to explain the observational data. In any case, before going to the strong-field regime $L\gg 1$ one has to check that at least one of the inequalities is violated. This can be done only numerically for each choice of the coupling function.

For spin-$1/2$ particles, as well as for the bosons in the region $M\ll L\lesssim 1$, the weak-field expression (\ref{weak-field-fermi}) could play a role only if
$e/(\pi f)\gtrsim 1$. However, this is never achieved during inflation, because $e/\pi<1$ and $f(\phi)>1$. Although the negative conductivity seems to be advantageous for the generation of an electric field, in the kinetic coupling model it has no effect. Therefore, for fermions, one has to consider only the strong-field regime.

The strong-field expressions for the scalar and spinor Schwinger currents are very similar and have the form
\begin{equation}
\label{schwinger-current-large-field}
j_{s}=\frac{a g_{s}}{12\pi^{3}}\frac{e^{3}E^{2}}{H}{\rm sign\,}(eE)e^{-\frac{\pi m^{2}}{|eE|}}, \quad s={b,\,f}.
\end{equation}
where $g_{b}=1$ and $g_{f}=2$ are the numbers of spin degrees of freedom. Expressing the Schwinger term in Eq.~(\ref{Maxwell-4}) in terms of energy density, we obtain
\begin{equation}
\label{strong-field}
\frac{1}{a}(j_{s}\cdot E)=\frac{g_{s}}{3\sqrt{2}\pi^{3}}\frac{e^{3}}{f^{3}}\frac{\rho_{E}^{3/2}}{H}e^{-\frac{f}{e}\frac{\pi m^{2}}{\sqrt{2\rho_{E}}}}.
\end{equation}
This term becomes comparable with $H\rho_{E}$ only at the late stages of inflation when
\begin{equation}
f\lesssim \frac{e}{\pi}\left(\frac{M_{p}}{H}\right)^{1/3}.
\end{equation}
In this case, the parameter $L$ is really large
\begin{equation}
L\sim \left(\frac{\pi f}{e}\right)^{2}\gg 1
\end{equation}
and expression (\ref{strong-field}) is applicable. 

Thus, the strong-field regime could become efficient only in the late stages of inflation, when the inflaton field [and the coupling function $f(\phi)$] changes very quickly. This results in an abrupt decrease of the electric energy density due to a high conductivity of produced plasma. Therefore, the main role of the Schwinger effect is to stop the generation of electromagnetic fields. After the effect turns on the magnetic field evolves according to the magnetic flux conservation law, i.e. $B\propto a^{-2}$. Therefore, the magnetic field power spectrum has to be determined before the Schwinger effect turns on.

It is important to mention that all the expressions for the Schwinger current were obtained in the case of a constant electric field energy density. However, such a behavior cannot be realized in an expanding Universe without some artificial currents which in addition could violate the second law of thermodynamics \cite{Giovannini:2018a}. Moreover, under the real circumstances, the electric field is time dependent and often grows very quickly. Nevertheless, the induced current due to Schwinger pair production was evaluated in Ref.~\cite{Kitamoto:2018} in the time-dependent electric background generated due to the kinetic coupling with the inflaton. The author considers the powerlike coupling function, $f\propto a^{\alpha}$, and pays particular attention to the case $\alpha=2$. The strong-field expression (4.21) in Ref.~\cite{Kitamoto:2018} for the induced current in the case $\alpha=2$ can be rewritten in our notations as follows:
\[
j_{b}=\frac{a}{28\pi^{3}}\frac{e^{3}E^{2}}{H}{\rm sign\,}(eE),
\]
which differs from our Eq.~(\ref{schwinger-current-large-field}) only by a factor of $3/7$ which is of the order unity. Therefore, our expression has the correct functional dependence and can be used for the numerical analysis of the Schwinger effect.

Finally, we also need an equation governing the evolution of the created particles. It is natural to require that the energy dissipated by electromagnetic fields be transfered into created particles.
Then, the energy density of produced particles can be described phenomenologically by the following equation:
\begin{equation}
\label{produced-density}
\dot{\rho}_{\chi}+4H\rho_{\chi}=\frac{1}{a}(j\cdot E),
\end{equation}
where the source term is given by Eq.~(\ref{strong-field}) [or by Eq.~(\ref{weak-field-bose}) if the weak-field regime for the scalar particles appears to be effective]. Since we consider the particles' masses smaller than the Hubble parameter, we treat the created particles as ultrarelativistic, which leads to the factor $4H$ in the above equation.

\section{Power spectra in the backreaction regime}
\label{sec-BR-spectra}

When the electric field energy density becomes large enough, the backreaction can change the regime of evolution of the inflaton field. Let us estimate the corresponding density when the backreaction starts to influence the background evolution. For this purpose, we consider Eq.~(\ref{KGF-2}), in which the right-hand side can be rewritten as follows:
\begin{equation}
\label{KGF-3}
\ddot{\phi}+3H\dot{\phi}+V'_{\phi}=2(\ln f)' \rho_{E}.
\end{equation}

We suppose that for a short time near the moment of the backreaction occurrence the coupling function behaves like a power of the scale factor $f\propto a^{-\gamma}$, $\gamma>0$ (we will see that just after entering the new regime of evolution  $\gamma\simeq 2$ independently of the explicit expression of the coupling function). Then we have
\begin{equation}
(\ln f)'=\frac{1}{\dot{\phi}}\frac{d \ln f}{dt}=-\gamma \frac{H}{\dot{\phi}}.
\end{equation}

Since inflation proceeds most of the time in the slow-roll regime, we can neglect the term with the second derivative in Eq.~(\ref{KGF-3}). Then we obtain the following relation:
\begin{equation}
\label{modified-slow-roll}
3H\dot{\phi}\left(1+\frac{2\gamma}{3}\frac{\rho_{E}}{\dot{\phi}^{2}}\right)+V'_{\phi}=0.
\end{equation}
It is obvious that the new regime appears when the second term in the brackets becomes of order unity, because in this case the time derivative $\dot{\phi}$ has to decrease in order to preserve the product (here we take into account that $V'_{\phi}$ as a function of $\phi$ but not $\dot{\phi}$ feels the changes later, when the field significantly deviates from its original trajectory). In other words, the backreaction becomes important when
\begin{equation}
\label{br-occurrence}
\rho_{E}\sim \frac{3}{2\gamma}\dot{\phi}^{2}\simeq \frac{V}{\gamma}\frac{M_{p}^{2}}{2}\left(\frac{V'_{\phi}}{V}\right)^{2}\sim \epsilon V\sim \epsilon \rho_{\rm inf}.
\end{equation}
Therefore, the backreaction occurs even earlier than the electric energy density becomes comparable with the inflaton energy density. As a result, the Friedmann equation (\ref{Friedmann}) does not feel directly the energy density of the generated field but only through the inflaton.

We would also like to mention that if the coupling function depended only on the curvature or scale factor but not on the inflaton field directly, there would not be the right-hand side in Eq.~(\ref{KGF-3}), and the only place where the electric energy density could cause the backreaction would be the Friedmann equation (\ref{Friedmann}). In this case, the backreaction would occur when the electric energy density became comparable with that of the inflaton, i.e. $\rho_{E}\sim \rho_{\rm inf}$. Firstly, the scale factor evolution would change in accordance with the Friedmann equation and after that the scalar field would feel it.

We would like to emphasize that Eq.~(\ref{br-occurrence}) was derived by using the two approximations: (i) the slow-roll regime (we neglected $\ddot{\phi}$) and (ii) that the electric energy density only begins to influence the evolution [the additional term in the parentheses in Eq.~(\ref{modified-slow-roll}) is of the order unity]. Therefore, Eq.~(\ref{br-occurrence}) can be used only to estimate the value of the electric energy density when the backreaction becomes important. During further evolution (especially at the end of inflation) the value of $\rho_{E}$ can be determined only numerically and is shown in Figs.~\ref{fig-evolution-power}(a) and \ref{fig-evolution-Ratra}(a) for the two particular choices of the coupling function.

Now, let us look at Eq.~(\ref{Maxwell-4}) without the Schwinger term. When the backreaction regime is established, the electric energy density becomes almost constant and its time derivative is small in comparison with the other terms, $\dot{\rho}_{E}\ll H\rho_{E}$. The boundary term on the right-hand side is important only at the early stages when the energy density is close to zero. When $\rho_{E}\sim \epsilon H^{2}M_{p}^{2}$, $H^{5}\ll H\rho_{E}$ for $H\ll M_{p}/\sqrt{\epsilon}$. Consequently, $\rho_E$ can be neglected. Then the two remaining terms must cancel out their leading contributions that give the condition
\begin{equation}
2H+\frac{\dot{f}}{f}\simeq 0,
\end{equation}
which immediately implies
\begin{equation}
a^{2}f={\rm const}
\end{equation}
independently of the initial time dependence of the coupling function.

This fact allows us to estimate the behavior of electric and magnetic power spectra for modes which cross the horizon after the backreaction regime becomes relevant. For such modes,
\begin{equation}
\dot{f}_{k}/f_{k}=2H_{k}, \quad f_{k}=\frac{C}{a_{k}^{2}}\propto k^{-2}, \quad 
\int_{t_{k}}^{t}\frac{a_{k}f^{2}_{k}}{a f^{2}}dt'=\frac{1}{3H}\left(\frac{a}{a_{k}}\right)^{3}\propto k^{-3}.
\end{equation}

Substituting these expressions into Eqs.~(\ref{E-power-spectrum-2})-(\ref{B-power-spectrum-2}), we obtain
\begin{equation}
\frac{d\rho_{E}}{d\ln k}\propto k^{0}, \quad \frac{d\rho_{B}}{d\ln k}\propto k^{2}.
\end{equation}

Since the backreaction regime once established lasts until the end of inflation (or until the Schwinger effect turns on), the last mode which crosses the horizon just before the end has no time to be enhanced. Therefore, we have the following expressions for the power spectra (up to a factor of order unity):
\begin{equation}
\label{spectra-br-regime}
\frac{d\rho_{E}}{d\ln k}=\frac{H^{4}}{4\pi^{2}},\quad \frac{d\rho_{B}}{d\ln k}=\frac{H^{4}}{4\pi^{2}}\left(\frac{k}{a_{e}H}\right)^{2}.
\end{equation} 

The present-day value of the observed magnetic field is determined by all relevant modes which can survive the further evolution of the Universe. Assuming the flux conservation, we have
\begin{equation}
\label{magnetic-field-today}
B_{0}=\left(\frac{a_{e}}{a_{0}}\right)^{2}\sqrt{2\int_{a_{i}H}^{k_{\rm diff}}\frac{dk}{k}\frac{d \rho_{B}}{d \ln \, k}},
\end{equation}
where $k_{\rm diff}$ is the momentum which now corresponds to the cosmic diffusion scale, i.e., the smallest size of magnetic configuration which can survive the cosmic diffusion in the late stages of the Universe's evolution. It could be estimated as $k_{\rm diff}/a_{0}\sim 1 \, {\rm A.U.}^{-1}=1.3\cdot 10^{-27}\,{\rm GeV}$ \cite{Grasso:2001}.

If the backreaction occurs well before $k_{\rm diff}$ crosses the horizon, then the magnetic power spectrum will have a blue tilt and the main contribution to the magnetic field comes from the upper integration bound
\begin{equation}
\label{analytical_B0}
B_{0}=\left(\frac{a_{e}}{a_{0}}\right)^{2}\frac{H^{2}}{2\pi}\left(\frac{k_{\rm diff}}{a_{e}H}\right).
\end{equation}

The value of $\frac{a_{0}}{a_{e}}$ can be found from the fact that the pivot scale $k_{\ast}$ crosses the horizon $N_{\ast}$ $e$-folds before the end of inflation:
\begin{equation}
\label{scale-factors-relation}
\frac{a_{0}}{a_{e}}=\frac{a_{\ast}}{a_{e}}\frac{a_{0}H_{\ast}}{k_{\ast}}=e^{-N_{\ast}}\frac{a_{0}H_{\ast}}{k_{\ast}}.
\end{equation}

Then the present-day strength of the magnetic field is given by the following expression:
\begin{eqnarray}
\label{B0}
B_{0}&=&\frac{M_{p}^{2}}{2\pi}\left(\frac{k_{\rm diff}}{a_{0}M_{p}}\right)\left(\frac{k_{\ast}}{a_{0}M_{p}}\right)e^{N_{\ast}}=\nonumber\\
&=&(1.6\cdot 10^{-23}\,{G})\ \left[\frac{1\,{\rm A.U.}}{r_{\rm diff}}\right] \left[\frac{500\,{\rm Mpc}}{\lambda_{\ast}}\right]e^{N_{\ast}-60}.
\end{eqnarray}

This equation sets the upper bound on the present value of the generated magnetic field in models where the backreaction regime is established well before the shortest relevant mode crosses the horizon. However, this estimate is incorrect in the case where the backreaction occurs after $k_{\rm diff}$ crosses the horizon. This situation will be considered numerically in the next section.

\section{Numerical results}
\label{sec-numerical}

In this section we analyze the process of electromagnetic field generation numerically. In order to be specific, we consider the Starobinsky model of inflation, which is favored by the latest results of the Planck Collaboration \cite{Planck:2018-infl}. The inflaton potential in this model has the form
\begin{equation}
V(\phi)=\frac{3}{4}\mu^{2}M_{p}^{2}\left(1-\exp\left[-\sqrt{\frac{2}{3}}\frac{\phi}{M_{p}}\right]\right)^{2},
\end{equation}
where the parameter $\mu$ can be fixed by the requirement that the amplitude of primordial scalar perturbations at the moment when the pivot mode $k_{\ast}$ crosses the horizon agrees with the CMB observations \cite{Planck:2018-infl} and equals
\begin{equation} 
\label{perturbation-amplitude}
\mathcal{P}_{\mathcal{R}}=\left.\left(\frac{H^{2}}{2\pi |\dot{\phi}|}\right)^{2}\right|_{N_{\ast}}=2.1\cdot 10^{-9}.
\end{equation}
From Eqs.~(\ref{Friedmann}) and (\ref{KGF-2}) in the slow-roll regime, taking $H^{2}=V/(3M_{p}^{2})$ we find
\begin{equation}
\dot{\phi}=-\frac{1}{3H}V'_{\phi}=-\sqrt{\frac{2}{3}}\mu M_{p}\exp\left(-\sqrt{\frac{2}{3}}\frac{\phi}{M_{p}}\right).
\end{equation}
Substituting it into Eq.~(\ref{perturbation-amplitude}), we find the value of the parameter $\mu$,
\begin{equation}
\frac{\mu}{M_{p}}=\frac{2\pi \sqrt{2\mathcal{P}_{\mathcal{R}}/3}}{{\rm sinh}^{2}\left(\frac{\phi_{\ast}}{\sqrt{6}M_{p}}\right)},
\end{equation}
where $\phi_{\ast}$ is the value of the inflaton field when the pivot scale $k_{\ast}$ crosses the horizon. Now, for fixed $k_{\ast}$ and $\mathcal{P}_{\mathcal{R}}$, the only parameter of the model is $\phi_{\ast}$. It can be found numerically from the requirement that the horizon crossing of the mode $k_{\ast}$ happens at $N_{\ast}$ $e$-folds before the end of inflation.

In order to solve the system of Eqs.~(\ref{Friedmann}), (\ref{KGF-2}) and (\ref{Maxwell-4}) numerically, we impose the following initial conditions:
\begin{equation}
\label{init-conditions}
a(0)=1,\quad \phi(0)=\phi_{0}, \quad \dot{\phi}(0)=-\frac{V'(\phi_{0})}{\sqrt{3V(\phi_{0})}}, \quad \rho_{E}(0)=\rho_{\chi}(0)=0,
\end{equation}
where $\phi_{0}$ has to be larger than $\phi_{\ast}$ because we wish to consider all physically relevant modes which, of course, include the pivot scale $k_{\ast}$.

The system of Eqs.~(\ref{Friedmann}), (\ref{KGF-2}), and (\ref{Maxwell-4}) determines how electric energy density evolves in time and does not allow one to extract the time evolution of the magnetic field which is the main purpose of our analysis. However, in the case $\rho_{B}\ll \rho_{E}$, it is possible to calculate the magnetic power spectrum perturbatively. Using the solution of Eqs.~(\ref{Friedmann}), (\ref{KGF-2}) and (\ref{Maxwell-4}) as a background, we numerically integrate Eq.~(\ref{mode-eq}) with the initial conditions (\ref{init-BD}) for all modes which cross the horizon during the inflation stage (until the moment when the Schwinger effect turns on). Then, using the mode functions, we calculate the magnetic field power spectrum (\ref{B-power-spectrum}) and compare it with the background electric energy density. If $\rho_{B}\ll \rho_{E}$, our approach is self-consistent. After that the present-day value of the magnetic field can be calculated using Eq.~(\ref{magnetic-field-today}).

We determine the time evolution of the magnetic energy density numerically by using Eq.~(\ref{mode-eq}), where the Schwinger effect is taken into account only indirectly through the evolution of the inflaton field. The latter is influenced by the electric field density whose dynamics is directly affected by the Schwinger term. A more self-consistent approach would require the inclusion of the Schwinger source term directly on the right-hand side of Eq.~(\ref{mode-eq}). However, including such a term is a nontrivial problem, because the Schwinger current depends on the total electric energy density and, therefore, contains contributions from all modes which undergo enhancement. Due to this fact all modes are coupled and there is no possibility to solve Eq.~(\ref{mode-eq}) for each separate mode.
	
There are some general physical arguments which can help to determine the behavior of the magnetic energy density. The Schwinger effect produces charged particles which form plasma. Due to the high conductivity of the latter the electric field significantly decreases and the magnetic field becomes ``frozen-in'' and starts to evolve in accordance with the flux conservation law, namely $B\propto a^{-2}$. We use this fact to calculate the present-day value of the magnetic field (\ref{magnetic-field-today}).

We consider the two types of coupling functions which are the most popular in the literature, the function which scales like $f\propto a^{\alpha}$ and the Ratra coupling function, $f=\exp(\beta\phi/M_{p})$. We would like to mention that for both coupling functions and relevant values of the parameters $\alpha$ and $\beta$ the infrared hyperconductivity in the bosonic Schwinger current plays no role for the masses $m> m_{\pi}\sim 0.1\,{\rm GeV}$; therefore, in our analysis we use the strong-field expressions for the Schwinger current (\ref{strong-field}).

\subsection{Coupling function $f\propto a^{\alpha}$}

Here we choose the coupling function which behaves as $f\propto a^{\alpha}$ during the slow-roll inflation. By using 
Eqs.~(\ref{Friedmann}) and (\ref{KGF-2}), we can express it in the slow-roll regime in terms of the inflaton field as follows:
\begin{equation}
\label{coupling-on-phi}
f(\phi)=\exp\left[\frac{3\alpha}{4}\left(1+\sqrt{\frac{2}{3}}\frac{\phi}{M_{p}}-e^{\sqrt{\frac{2}{3}}\frac{\phi}{M_{p}}}\right)\right].
\end{equation}
Since the coupling function primarily depends on the inflaton field $\phi$ rather than on the scale factor and it is difficult to think of the mechanism which ensures scaling $\propto a^{\alpha}$, we will consider expression (\ref{coupling-on-phi}) as a definition that gives the scaling $f\propto a^{\alpha}$ in the slow-roll regime.

It is instructive to verify Eq.~(\ref{Maxwell-4}) analytically for the above coupling function in de Sitter space-time with constant $H$. 
Martin and Yokoyama showed \cite{Martin:2008} that the electric field contribution dominates over the magnetic field contribution in this case 
and its power spectrum is given by
\begin{equation}
\frac{d\rho_{E}}{d\ln\,k}=\frac{H^{4}}{2\pi^{2}}G(\alpha)\left(\frac{k}{aH}\right)^{2\alpha+4}, \quad G(\alpha)=\frac{\pi}{2^{2\alpha+3}\Gamma^{2}(\alpha+3/2)\cos^{2}(\pi\alpha)}.
\end{equation}

Then the total energy density of the superhorizon modes equals
\begin{equation}
\label{rhoE-alpha}
\rho_{E}=\int_{H}^{aH}\frac{dk}{k}\frac{d\rho_{E}}{d\ln\,k}=\frac{H^{4}}{2\pi^{2}}\frac{G(\alpha)}{(2\alpha+4)}\left(1-a^{-(2\alpha+4)}\right).
\end{equation}

Substituting (\ref{rhoE-alpha}) into the left-hand side of Eq.~(\ref{Maxwell-4}), we obtain
\begin{equation}
\label{exact}
\dot{\rho}_{E}+4H\rho_{E}+2\frac{\dot{f}}{f}\rho_{E}=\frac{H^{5}}{2\pi^{2}}G(\alpha).
\end{equation}
In the absence of charged fields, the right-hand side for $f\sim a^{\alpha}$ reads
\begin{equation}
\frac{H^{5}}{4\pi^{2}}(1+\alpha^{2}).
\end{equation}
We see that they have similar parametric behavior and differ only by a factor of order unity. Therefore, the approximate expression (\ref{boundary}) is acceptable.

\begin{figure}[h!]
	\centering
	\includegraphics[width=0.49\textwidth]{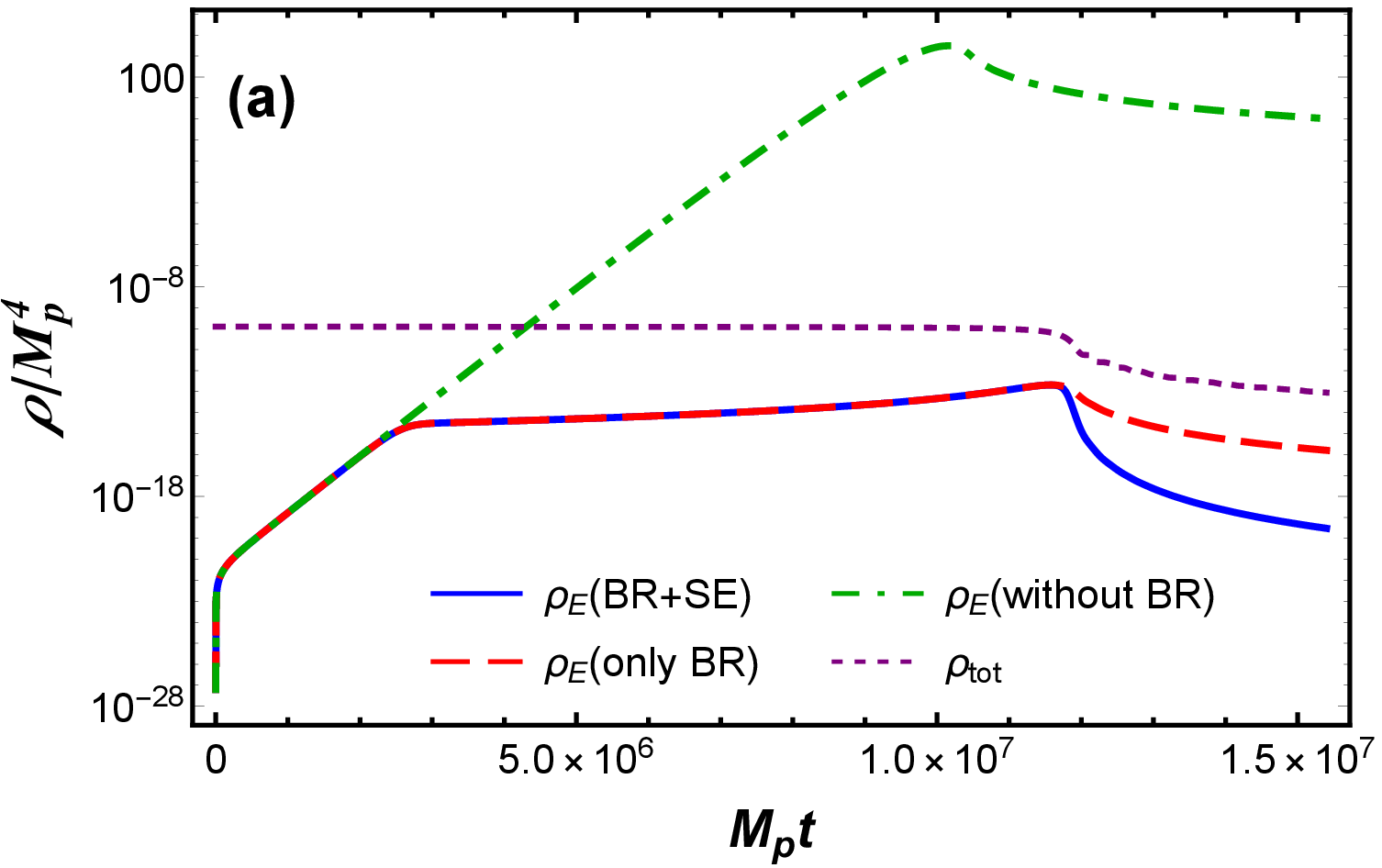}
	\includegraphics[width=0.47\textwidth]{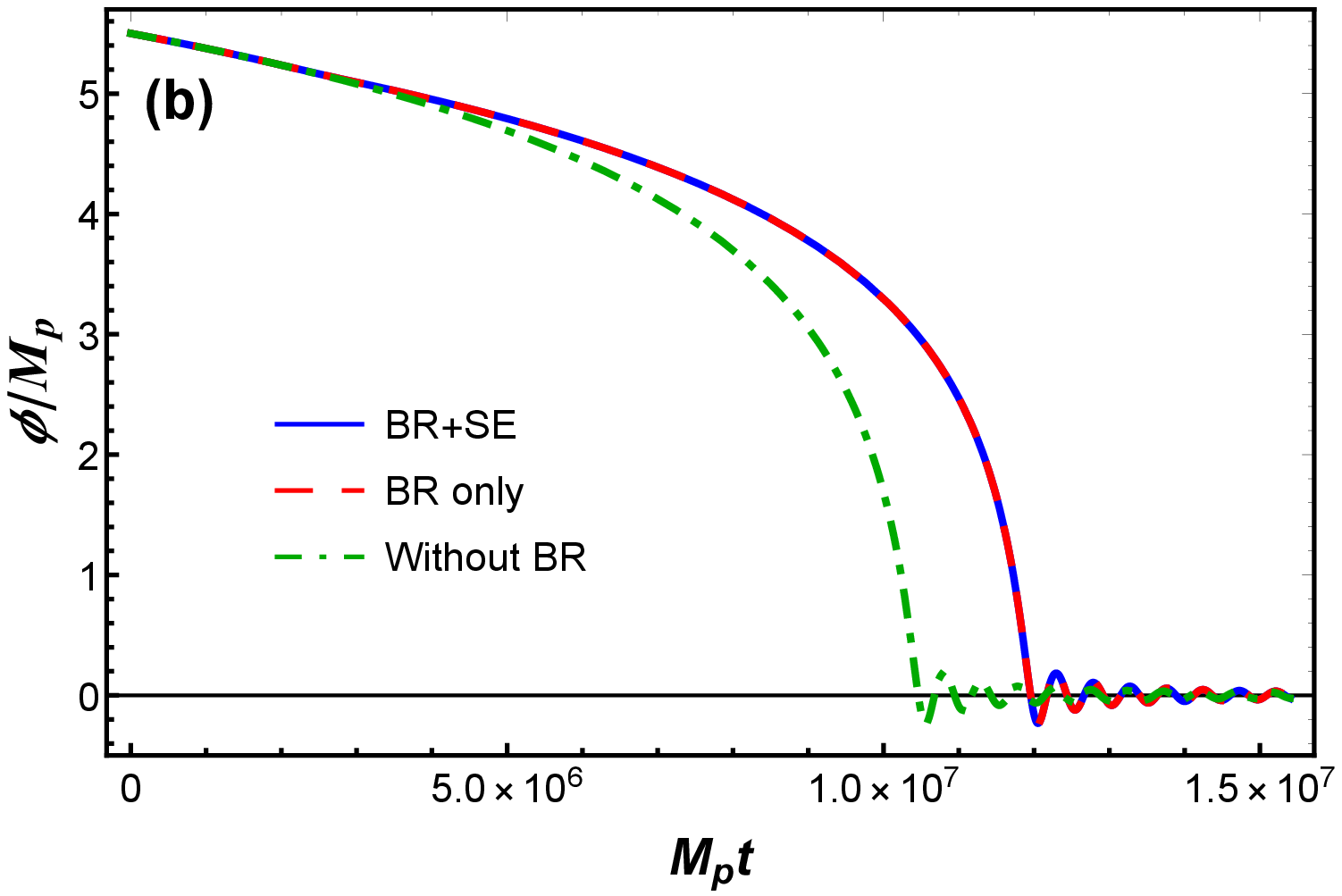}
	\caption{The time dependence of the electric energy density (a) and the inflaton field (b) for the powerlike coupling function with
	$\alpha=-2.5$.  The green dashed-dotted lines show the case where the backreaction and Schwinger effect are not taken into account; the red dashed lines 
	correspond to the case when the backreaction is included and the Schwinger effect is absent. Finally, the blue solid lines take into account 
	the Schwinger effect. The purple dotted line in panel (a) shows the total energy density $\rho_{\rm tot}=3H^{2}M_{p}^{2}$.
	\label{fig-evolution-power}}
\end{figure}

Further, we will perform a numerical analysis of the system. We set $\alpha=-2.5$ (so that the backreaction problem really occurs) and show in Fig.~\ref{fig-evolution-power} the time dependencies of the electric field energy density and the inflaton field. At first, we consider only the situation when the backreaction from electric field is present in Eqs.~(\ref{Friedmann}) and (\ref{KGF-2}) and the Schwinger effect is absent (see the red dashed lines in Fig.~\ref{fig-evolution-power}). In other words, this corresponds to the situation when there are no charged fields in the Universe.

It is obvious that at low energy density the evolution is the same as for the case without backreaction. Since the coupling function for $\alpha<-2$ decreases faster than $a^{2}$ all the time, the third term on the left-hand side of Eq.~(\ref{Maxwell-4}), which is negative, is always stronger than the second one, which is positive. Hence their interplay leads to the enhancement of the electric field energy density like $\propto a^{2|\alpha|-4}$; see Eq.~(\ref{rhoE-alpha}). Each separate mode starts to grow just after it crossed the horizon and the earlier it crossed, the larger amplitude it will have. This leads to a red-tilted power spectrum in the regime without backreaction, reported in Ref.~\cite{Martin:2008} and which can be seen in Fig.~\ref{fig-spectrum-BR}. However, a different behavior arises when the electric field becomes strong enough to affect the background evolution equations. In this case it slows down the rolling of the scalar field. Also, as it was shown in Sec.~\ref{sec-BR-spectra}, the coupling function evolves approximately like $\propto a^{-2}$ and the two terms in Eq.~(\ref{Maxwell-4}) almost compensate each other. In this regime, the electric field energy density becomes almost constant until the end of inflation. The backreaction prolongs also inflation.

\begin{figure}[h!]
	\centering
	\includegraphics[width=0.49\textwidth]{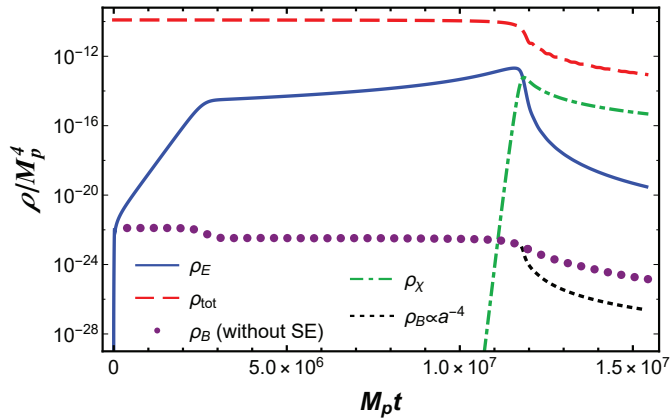}
	\caption{The time dependence of the electric energy density (blue solid line), the total energy density (red dashed line), and the energy density of the charged particles  generated due to the Schwinger process (green dashed-dotted line) for a powerlike coupling function with $\alpha=-2.5$. The time dependence of the magnetic energy density calculated perturbatively is shown by the purple dots (without taking into account the Schwinger effect). The black dotted line shows the adiabatic evolution of the magnetic energy density, $\rho_{B}\propto a^{-4}$, after the Schwinger effect is turned on.
		\label{fig-magnetic-BR}}
\end{figure}

Finally, we take into account the Schwinger effect. In numerical calculations we use the strong-field expression for the Schwinger source term given by Eq.~(\ref{strong-field}). Since it is proportional to the cube of effective charge $e/f$, the Schwinger mechanism is inefficient during the early stages when $f$ is very large. However, it becomes relevant at the late stages and then strongly diminishes the electric field density [see the blue solid lines in Figs.~\ref{fig-evolution-power}(a) and \ref{fig-magnetic-BR}].

The energy density of the charged particles created due to the Schwinger process also becomes non-negligible only in the late stages. We plot its time dependence in Fig.~\ref{fig-magnetic-BR} by the green dashed-dotted line. After the Schwinger effect turns on it quickly transfers almost all the energy density from the electric field to the charged particles, starting the process of reheating even before the fast oscillations of the inflaton field (the so-called ``Schwinger reheating'' mentioned in Ref.~\cite{Tangarife:2017}). It is important to note that the expression for the Schwinger current (\ref{strong-field}) was derived in de Sitter space-time; therefore, its application during the stage of preheating is not well justified.

Finally, we carry out the perturbative calculation of the magnetic energy density. Using the background solutions for the inflaton and scale factor we numerically solve the mode equation (\ref{mode-eq}) for all modes which crossed the horizon during inflation. Calculating the magnetic energy density at a given moment of time we include only the modes which are outside the horizon at that moment. The corresponding dependence is shown in Fig.~\ref{fig-magnetic-BR} by the purple dots. These results confirm the applicability of the perturbative approach because the magnetic energy density is indeed much smaller than the electric one. The time evolution of the magnetic energy density during the preheating stage in the absence of the Schwinger effect is shown by purple dots in Fig.~\ref{fig-magnetic-BR}. It should be compared with the $\rho_{B}\propto a^{-4}$ behavior plotted by the black dotted line which takes place when the Schwinger effect is present.

\begin{figure}[h!]
	\centering
	\includegraphics[width=0.49\textwidth]{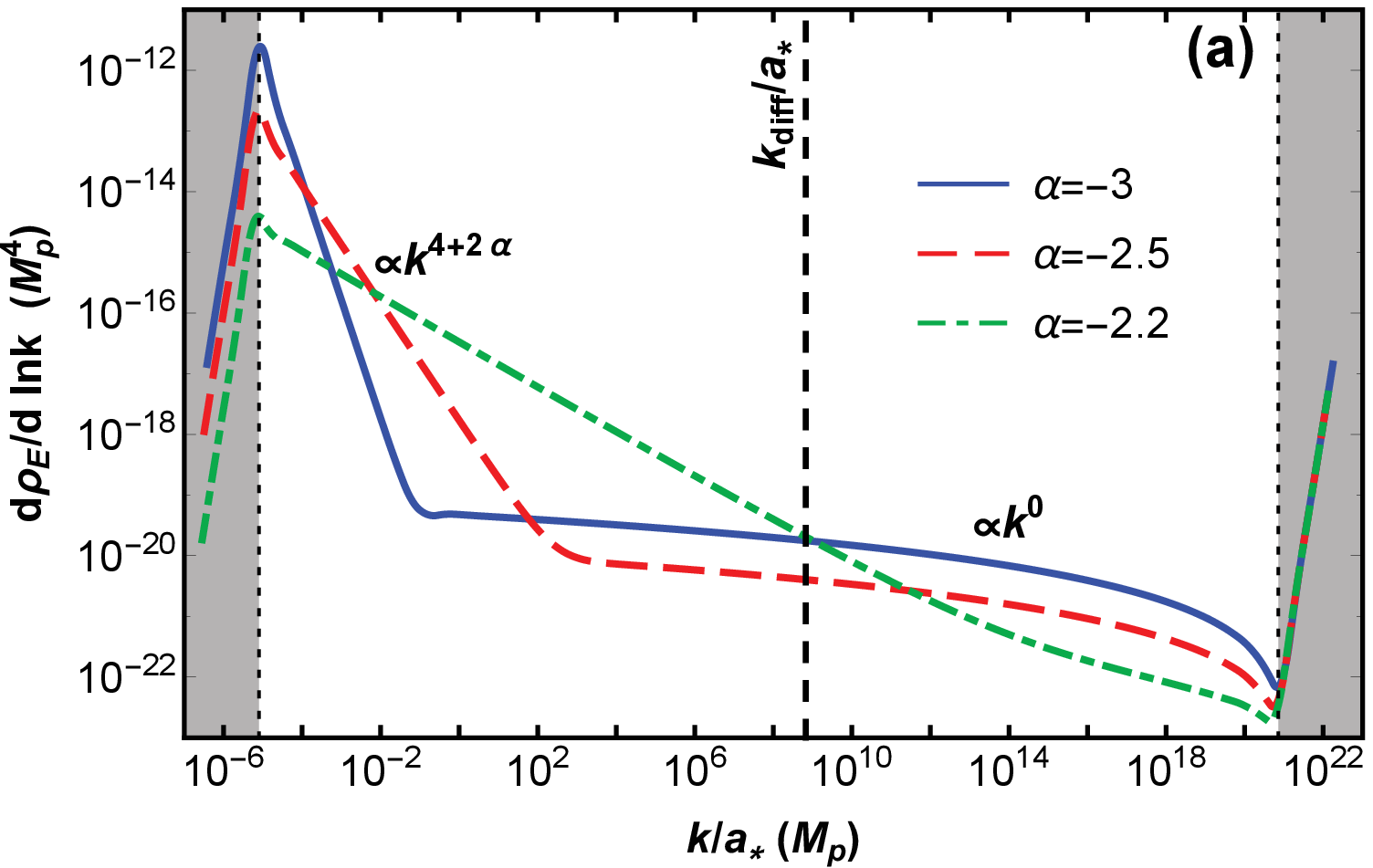}
	\includegraphics[width=0.49\textwidth]{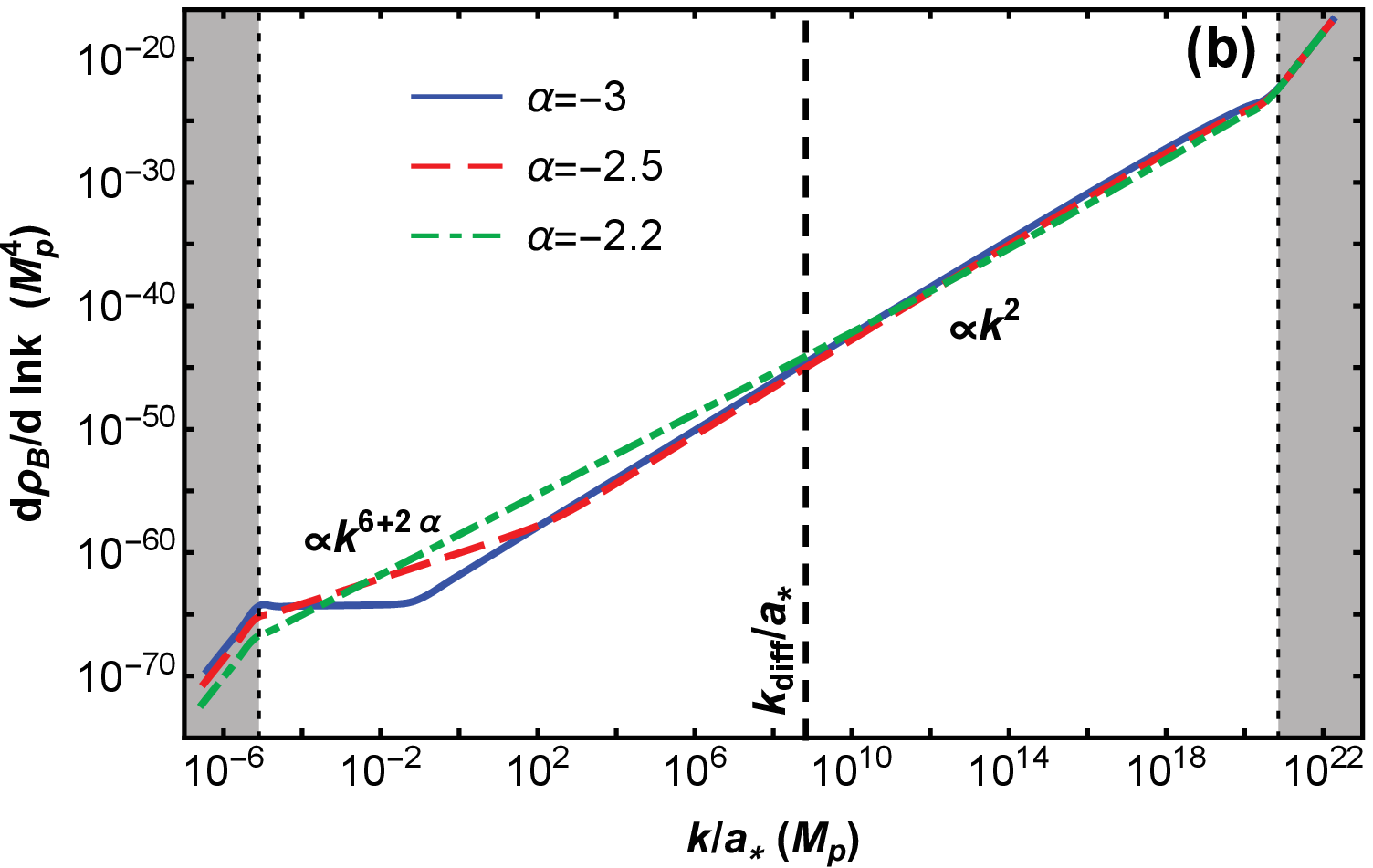}
	\caption{The electric (a) and magnetic (b) power spectra in the model with powerlike coupling function $f\propto a^{\alpha}$ for $\alpha=-2.2,\ -2.5$, and $-3$. Left shaded regions correspond to the modes which are outside the horizon even before the beginning of inflation. Right shaded regions correspond to the modes which have not crossed the horizon even until the end of inflation and, hence, do not undergo enhancement. The vertical black dashed lines show the momentum which corresponds to the cosmic diffusion scale $k_{\rm diff}/a_{\ast}$.
	 \label{fig-spectrum-BR}}
\end{figure}

Then, we analyze the power spectra of generated fields and their dependence on the parameter $\alpha$. Figure~\ref{fig-spectrum-BR} shows the electric [panel (a)] and magnetic [panel (b)] power spectra for three different values of the power $\alpha=-2.2,\ -2.5$, and $-3$ at the time when the Schwinger effect becomes efficient and inflation ends. We see that for modes which cross the horizon before the backreaction becomes relevant the power spectrum behaves like in the case where the backreaction is absent, i.e. $d\rho_{E}/d \ln k\propto k^{4+2\alpha}$ and $d\rho_{B}/d \ln k\propto k^{6+2\alpha}$ \cite{Martin:2008}. However, for shorter modes, which cross the horizon after the setting of the backreaction regime, the power spectra have similar scaling, independent of $\alpha$, very close to that predicted in Eq. (\ref{spectra-br-regime}).

\begin{figure}[h!]
	\centering
	\includegraphics[width=0.49\textwidth]{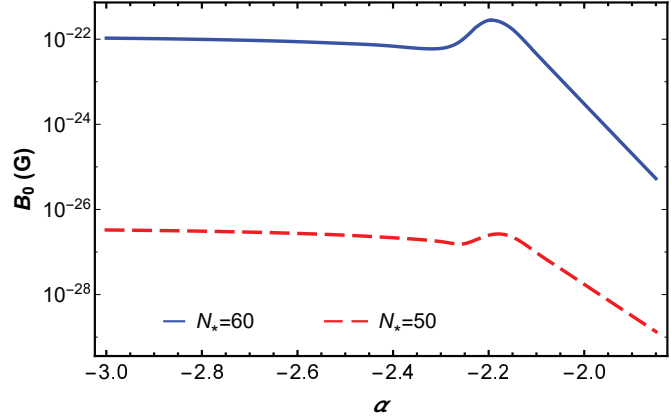}
	\caption{The dependence of the present-day value of the magnetic field on parameter $\alpha$ for two numbers of $e$-folds from the moment of the pivot scale horizon crossing to the end of inflation, $N_{\ast}=50$ (red dashed curve) and $N_{\ast}=60$ (blue solid curve).
		\label{fig-magnetic-today-power}}
\end{figure}

Having determined the magnetic power spectrum we can calculate the value of the magnetic field at the present epoch using Eq.~(\ref{magnetic-field-today}). We have to include all modes which crossed the horizon after the beginning of inflation and which are longer than the cosmic diffusion scale that is estimated today as $r_{\rm diff}/a_{0}\sim 1\,{\rm A.U.}$ \cite{Grasso:2001}. The dependence of the present-day value of the magnetic field on the parameter $\alpha$ is shown in Fig.~\ref{fig-magnetic-today-power} for the two values of the number of $e$-folds at which the pivot scale crosses the horizon, $N_{\ast}=50$ (red dashed line) and $N_{\ast}=60$ (blue solid line). Since we know the value of the pivot scale today and we can calculate its value at the moment of horizon crossing in terms of the Hubble parameter at that moment, we can find the total number of $e$-folds $N_{\rm tot}$ between these two moments of time. Fixing the number of $e$-folds from the pivot scale horizon crossing to the end of inflation $N_{\ast}$ we automatically fix the number of $e$-folds from the end of inflation to today; see Eq.~(\ref{scale-factors-relation}). Taking into account that the postinflationary evolution of the magnetic field is approximately determined by the flux conservation, $B\sim a^{-2}$, and that its spectral index is $n_{B}<4$, we conclude that the smaller $N_{\ast}$ we take, the smaller the value of the magnetic field we obtain, which is obvious also from Fig.~\ref{fig-magnetic-today-power}.

Since the magnetic power spectrum is blue [see Fig.~\ref{fig-spectrum-BR}(b)], the main contribution to the present-day value of the magnetic field $B_{0}$ is due to the modes with $k\lesssim k_{\rm diff}$. For $\alpha>-2$ the backreaction never occurs and the magnetic power spectrum behaves like $d\rho_{B}/d \ln k\propto k^{6+2\alpha}$; see e.g. Ref.~\cite{Martin:2008}. Its spectral index, $n_{B}=6+2\alpha>2$, increases with the increase of $\alpha$ and the amplitude of the spectrum has to decrease in order to match the unperturbed power spectrum for the modes with $k>k_{\rm max}$. As a result, the present-day value of the magnetic field is dependent on $\alpha$ and falls when $\alpha$ increases. However, for $\alpha<-2$ the backreaction starts to play an important role and modifies the spectrum; see Fig.~\ref{fig-spectrum-BR}. In particular, for $\alpha\lesssim-2.2$ it changes the spectrum for the modes, which are longer than the cosmic diffusion scale, and influences the present-day value of the magnetic field. For lower values of $\alpha$ it becomes independent of $\alpha$ and its value is in accordance with the upper value (\ref{B0}). The smaller-than-one-order-of-magnitude discrepancy between the analytically derived limit (\ref{B0}) and the exact numerical value can be explained by the time dependence of the Hubble parameter close to the end of inflation and by the slight deviation of the spectral index from $n_{B}=2$. For $-2.2<\alpha<-2$, the backreaction becomes important only when the modes with $k>k_{\rm diff}$ exit the horizon. As a result, for all the modes which contribute to $B_{0}$, the spectrum is not described by Eq.~(\ref{spectra-br-regime}) and has the spectral index $n_{B}=2\alpha+6< 2$; i.e., their spectrum is less steep than for $\alpha<-2.2$. At the same time, the amplitude of the spectrum is larger [see the green dashed-dotted line in Fig.~\ref{fig-spectrum-BR}(b) which is less steep and is located higher than the red dashed and blue solid lines]. This leads to larger values of $B_{0}$ for $-2.2<\alpha<-2$ compared to those for $\alpha<-2.2$. This can be seen as a bump in Fig.~\ref{fig-magnetic-today-power}.

As it was mentioned above, another feature of the generated magnetic field is its blue spectrum. This leads to very small coherence scales, comparable to $1$~A.U. at the present time. Since $n_{B}\approx2$, the values of the large-scale magnetic field on the Mpc scale are then $11$ orders of magnitude smaller. In any case, the value of the magnetic field is too small to be in accordance with observational data \cite{Neronov:2010,Tavecchio:2010,Taylor:2011,Dermer:2011,Caprini:2015}.

It is important to mention the problem of the models with a red-tilted power spectrum, which, in particular, we have for $\alpha<-2$. For such models it is very important when exactly the inflation starts, because all modes which cross the horizon during inflation undergo enhancement and it is larger for modes which cross the horizon earlier. For example, the power spectra in Fig.~\ref{fig-spectrum-BR} and the present-day values of the magnetic field in Fig.~\ref{fig-magnetic-today-power} were calculated with the assumption that the inflation lasts only $N_{e}=60$ $e$-folds. However, if we assume that it started much earlier, this will require us to include more modes from the infrared region and will lead to earlier setting of the backreaction regime. As a result, for all modes of physical relevance (which crossed the horizon 50--60 $e$-folds before the end of inflation) the evolution will occur in the backreaction regime and their power spectra will be given by Eq.~(\ref{spectra-br-regime}). In other words, if the inflation lasts much more than 60 $e$-folds, the power spectra of the generated electromagnetic fields for arbitrary $\alpha<-2$ will be equivalent to the case $\alpha=-2$, because the electric power spectrum will be scale-invariant and the magnetic one will have the spectral index $n_{B}\approx 2$. In this situation the principal upper bound on the present-day magnetic field (\ref{B0}) is valid.

Our results are in accordance with Ref.~\cite{Kanno:2009}, where the backreaction of the electric fields was taken into account for the powerlike coupling function. We confirm that for all $\alpha<-2$ in the backreaction regime the power spectra are equivalent to the ``attractor'' case $\alpha=-2$ and this strongly suppresses the generation of magnetic fields.

\subsection{Ratra coupling function $f=\exp(\beta \phi/M_{p})$}

Another example is given by the coupling function of the Ratra type
\begin{equation}
f(\phi)=\exp(\beta\phi/M_{p}).
\label{Ratra-coupling}
\end{equation}
Magnetogenesis in the Starobinsky model with the kinetic coupling (\ref{Ratra-coupling}) was studied numerically in Ref.~\cite{Vilchinskii:2017}. However, not all relevant modes were taken into account and the backreaction was not correctly estimated. Here we determine the electric and magnetic 
power spectra in this model taking into account the backreaction and Schwinger effect.

In the slow-roll regime and in the absence of backreaction, one can derive the following approximate expressions for the time dependence of the
scale factor and the inflaton field \cite{Vilchinskii:2017}:
\begin{eqnarray}
\label{scale-factor}
a(t)&=&\exp\left(\frac{\mu t}{2}\right)\left[1-\frac{t}{t_{e}}\right]^{3/4},\\
\label{inflaton-field}
\phi(t)&=&\phi_{0}+M_{p}\sqrt{\frac{3}{2}}\ln\left[1-\frac{t}{t_{e}}\right],
\end{eqnarray}
where $t_{e}=3e^{\sqrt{2/3}(\phi_{0}/M_{p})}/(2\mu)$ is the moment of time close to the end of inflation when the slow-roll approximation becomes
inapplicable. Equation (\ref{scale-factor}) makes it possible to define the moment of time $t_{k}$ when the mode with comoving momentum $k$
crosses the horizon
\begin{equation}
t_{k}\simeq\frac{2}{\mu}\ln\frac{2k}{\mu}.
\end{equation}
This expression is more accurate for long-wavelength modes, for which $t_{k}\ll t_{e}=\frac{2}{\mu}N_{e}$, $N_{e}=\frac{3}{4}e^{\sqrt{\frac{2}{3}}\frac{\phi_{0}}{M_{p}}}$. Then, using Eq.(\ref{inflaton-field}), we can calculate $f_{k}$:
\begin{equation}
f_{k}=f(t_{k})\simeq e^{\beta\phi_{0}/M_{p}}\left(1-\frac{t_{k}}{t_{e}}\right)^{\sqrt{\frac{3}{2}}\beta}\approx e^{\beta\phi_{0}/M_{p}}e^{-\frac{t_{k}}{t_{e}}\sqrt{\frac{3}{2}}\beta}=e^{\beta\phi_{0}/M_{p}}\left(\frac{2k}{\mu}\right)^{-\sqrt{\frac{3}{2}}\frac{\beta}{N_{e}}}.
\end{equation}
In addition,
\begin{equation}
\left|\frac{1}{H_{k}}\frac{\dot{f}_{k}}{f_{k}}\right|\simeq \frac{2\sqrt{2/3}\beta e^{-\sqrt{2/3}\phi_{0}/M_{p}}}{1-\frac{t_{k}}{t_{e}}}\lesssim 1.
\end{equation}

Therefore, using Eq.(\ref{E-power-spectrum-2}), the electric power spectrum reads (setting $f=1$ at the end of inflation)
\begin{equation}
\label{E-spectrum-Ratra}
\frac{d\rho_{E}}{d\ln\, k}\simeq\frac{k^{4}}{4\pi^{2}a_{e}^{4}}f_{k}^{2}\simeq\frac{H^{4}}{4\pi^{2}}\left(\frac{4N_{e}}{3e}\right)^{\sqrt{6}\beta}\left(\frac{k}{a_{e}H}\right)^{4-\sqrt{6}\frac{\beta}{N_{e}}}\propto k^{2+s},
\end{equation}
where $s=2-\sqrt{6}\frac{\beta}{N_{e}}$ is the so-called ``anomalous slope,'' introduced in Ref.~\cite{Vilchinskii:2017}. For $N_{e}=60$,
we have $s=2-0.04\beta$, which is very close to the result found numerically in Ref.~\cite{Vilchinskii:2017}.

To estimate the magnetic power spectrum we take into account that near the end of inflation $a f^{2}$ decreases very quickly and the the term proportional to $C_{2}$ in Eq.~(\ref{B-power-spectrum-2}) dominates. The corresponding integral can be estimated as
\begin{equation}
\int_{t_{k}}^{t}\frac{dt'}{a f^{2}}\sim \frac{1}{aH f^{2}}.
\end{equation}
Then,
\begin{equation}
\label{B-spectrum-Ratra}
\frac{d\rho_{B}}{d\ln\, k}\simeq\left(\frac{k}{a_{e}H}\right)^{2}\frac{d\rho_{E}}{d\ln\, k}\propto k^{4+s},
\end{equation}
which also is in agreement with Ref.~\cite{Vilchinskii:2017}. 

However, the power spectra (\ref{E-spectrum-Ratra}) and (\ref{B-spectrum-Ratra}) are only applicable for relatively low momenta $k\ll a_{e}H$ and for shorter modes more accurate expressions for $f_{k}$, $t_{k}$ must be used. Moreover, when the strong backreaction regime takes place the power spectra start to behave like that defined in Eq.(\ref{spectra-br-regime}).

\begin{figure}[h!]
	\centering
	\includegraphics[width=0.49\textwidth]{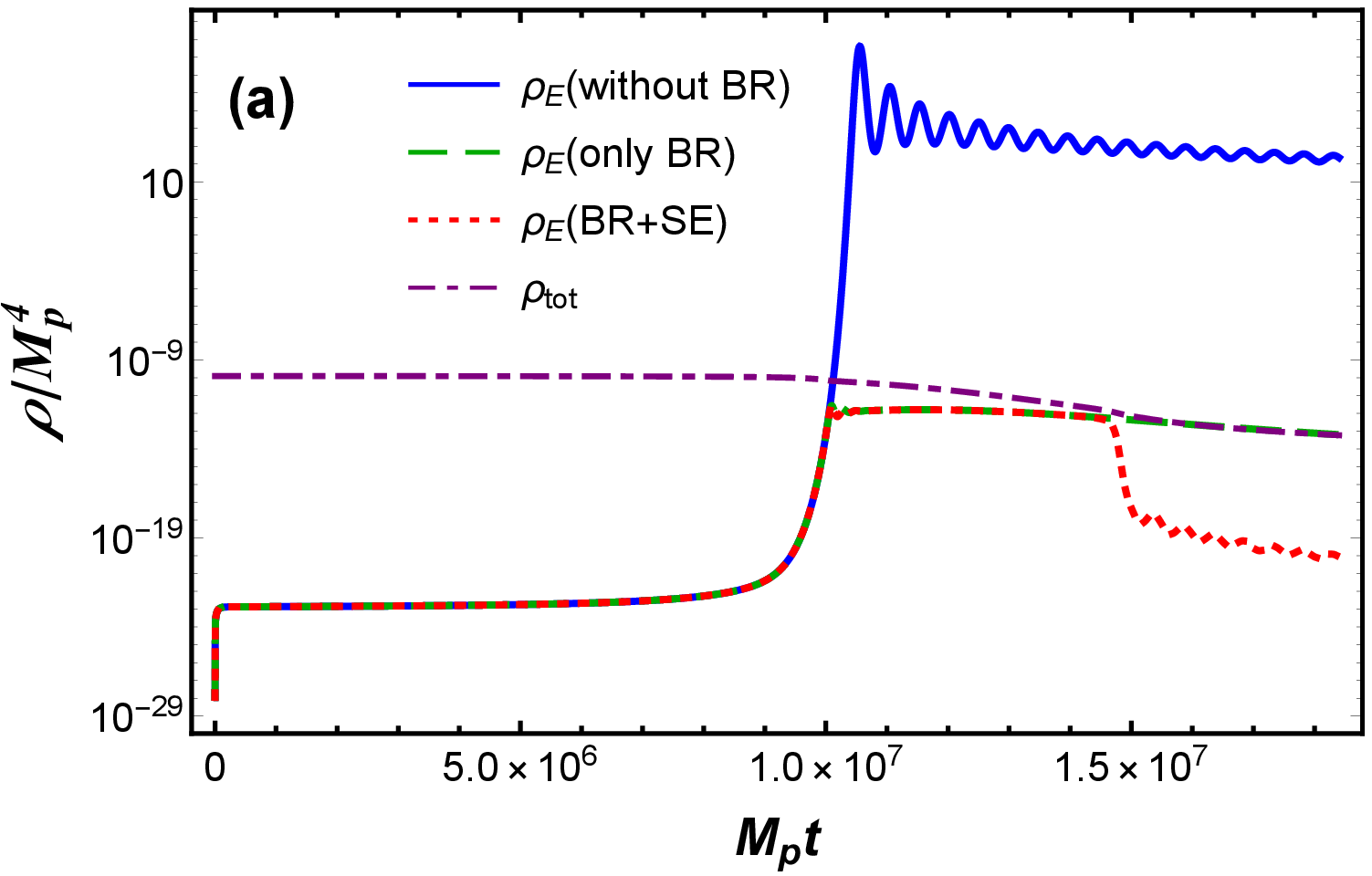}
	\includegraphics[width=0.47\textwidth]{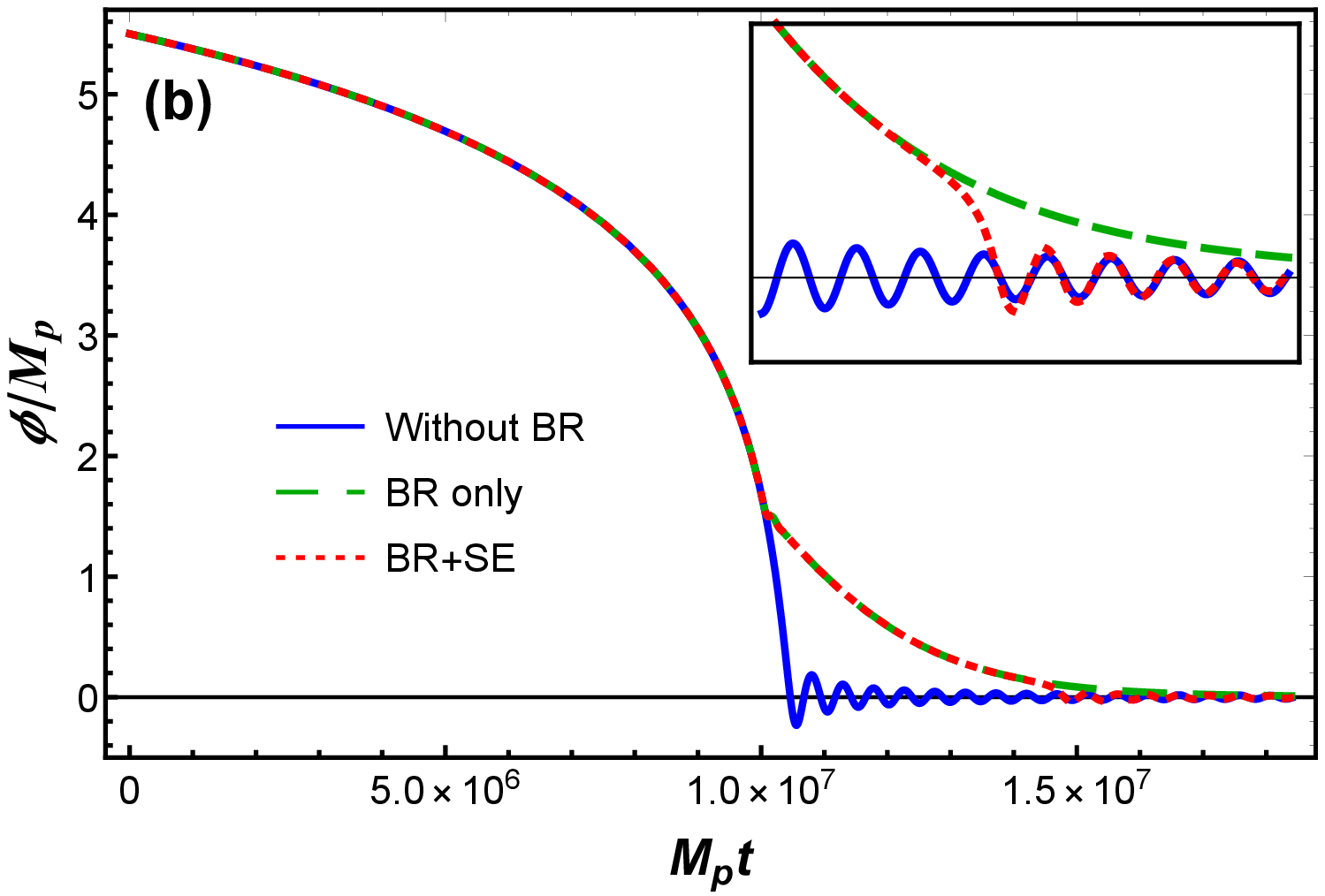}
	\caption{The time dependence of the electric energy density (a) and the inflaton field (b) for the Ratra coupling function with $\beta=15$.  The blue solid lines show the case when the backreaction and the Schwinger effect are not taken into account; the green dashed lines correspond to the case when the backreaction is included and the Schwinger effect is absent. Finally, the red dotted lines take into account the Schwinger effect. The purple dashed-dotted line shows the time dependence of the total energy density $\rho_{\rm tot}=3H^{2}M_{p}^{2}$.		\label{fig-evolution-Ratra}}
\end{figure}

In our numerical analysis, we use $\beta=15$ so that the backreaction problem indeed takes place. Figure~\ref{fig-evolution-Ratra} shows the time dependence of the energy density of the electric field [panel (a)] and the inflaton field [panel (b)]. The green dashed lines correspond to the situation when the back reaction is taken into account and the Schwinger effect is switched off. For comparison we also plot the corresponding dependencies for the case when the electromagnetic field is a spectator field (i.e., it feels the background but does not backreact). 

At the early stages of inflation the inflaton field is in the slow-roll regime and the coupling function also changes slowly; namely, it behaves like a power function [see Eq.~(\ref{inflaton-field})]. Under  such circumstances, the Hubble term in Eq.~(\ref{Maxwell-4}) damps the evolution of the electric field energy density. It would tend to zero like $a^{-4}$ if there were no boundary term on the right-hand side, which describes the new modes which cross the horizon and give their contribution to the electric energy density. Its presence leads to a dynamical equilibrium with electric energy density $\rho_{E}\sim H^{4}/(16\pi^{2})$ being almost constant at early times. 

At a certain moment of time, the negative term $2(\dot{f}/f) \rho_{E}$ becomes larger in the absolute value than the positive Hubble term $4H\rho_{E}$ and the electric energy density starts growing. As it was shown in Sec.~\ref{sec-BR-spectra}, when it reaches $\rho_{E,{\rm cr}}\sim \epsilon \rho_{\rm inf}$, the backreaction becomes very important and a new regime of evolution is established. The requirement that $f\propto a^{-2}$ in this regime significantly slows down the inflaton field and it deviates from its original trajectory. Figure~\ref{fig-evolution-Ratra} shows that the backreaction changes drastically the end of inflation, making the inflaton slide towards the minimum at $\phi=0$ without oscillations. This means the absence of the preheating stage in this case. 

\begin{figure}[h!]
	\centering
	\includegraphics[width=0.49\textwidth]{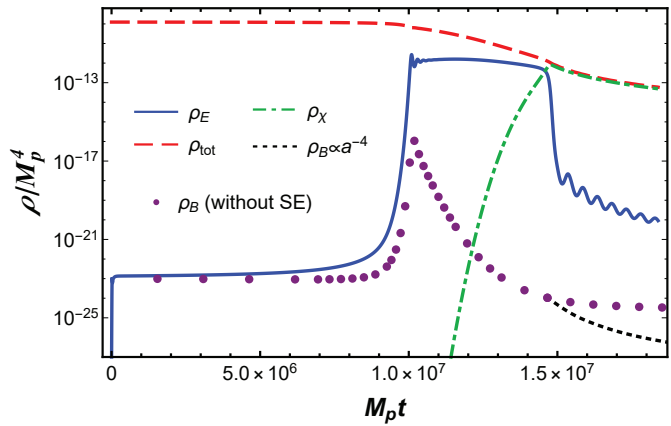}
	\caption{The time dependence of the electric energy density (blue solid line), the total energy density (red dashed line), and the energy density of the charged particles  generated due to the Schwinger process (green dashed-dotted line) for a Ratra coupling function with $\beta=15$. The time dependence of the magnetic energy density calculated perturbatively is shown by the purple dots (without taking into account the Schwinger effect). The black dotted line shows the adiabatic evolution of the magnetic energy density, $\rho_{B}\propto a^{-4}$, after the Schwinger effect is turned on.
	\label{fig-magnetic-Ratra}}
\end{figure}

Taking into account the Schwinger effect resolves this problem (see the red dotted lines in Fig.~\ref{fig-evolution-Ratra}). When the Schwinger current becomes significant it quickly diminishes the electric field and its backreaction. Then the inflaton field quickly rolls down to the potential minimum and oscillates. This helps us to restore the preheating stage, during which different particles can be created due to the parametric resonance processes \cite{Kofman:1997}. In addition, the Schwinger effect by itself fills the Universe with created charged particles, implementing the Schwinger reheating scenario. This can be seen from Fig.~\ref{fig-magnetic-Ratra}, where the energy density of created particles is shown by the green dashed-dotted line.

At the same time, we should note that the expression for the Schwinger current (\ref{strong-field}) was derived in de Sitter space-time. Therefore, its application during the stage of preheating is not well justified.

The magnetic field energy density can be calculated perturbatively. Figure~\ref{fig-magnetic-Ratra} shows its time dependence in comparison with 
the electric energy density. The results confirm the applicability of the perturbative approach as the magnetic energy density is indeed much 
smaller than the electric one. The figure shows clearly that after the backreaction becomes relevant the magnetic field begins to decrease in time. Purple dots show the time evolution of the magnetic field energy density in the absence of the Schwinger effect. However, if the latter is present, then the high conductivity of produced plasma leads to the adiabatic evolution of the magnetic field for which $\rho_{B}\propto a^{-4}$. This is shown in Fig.~\ref{fig-magnetic-Ratra} by the black dotted line.

\begin{figure}[h!]
	\centering
	\includegraphics[width=0.49\textwidth]{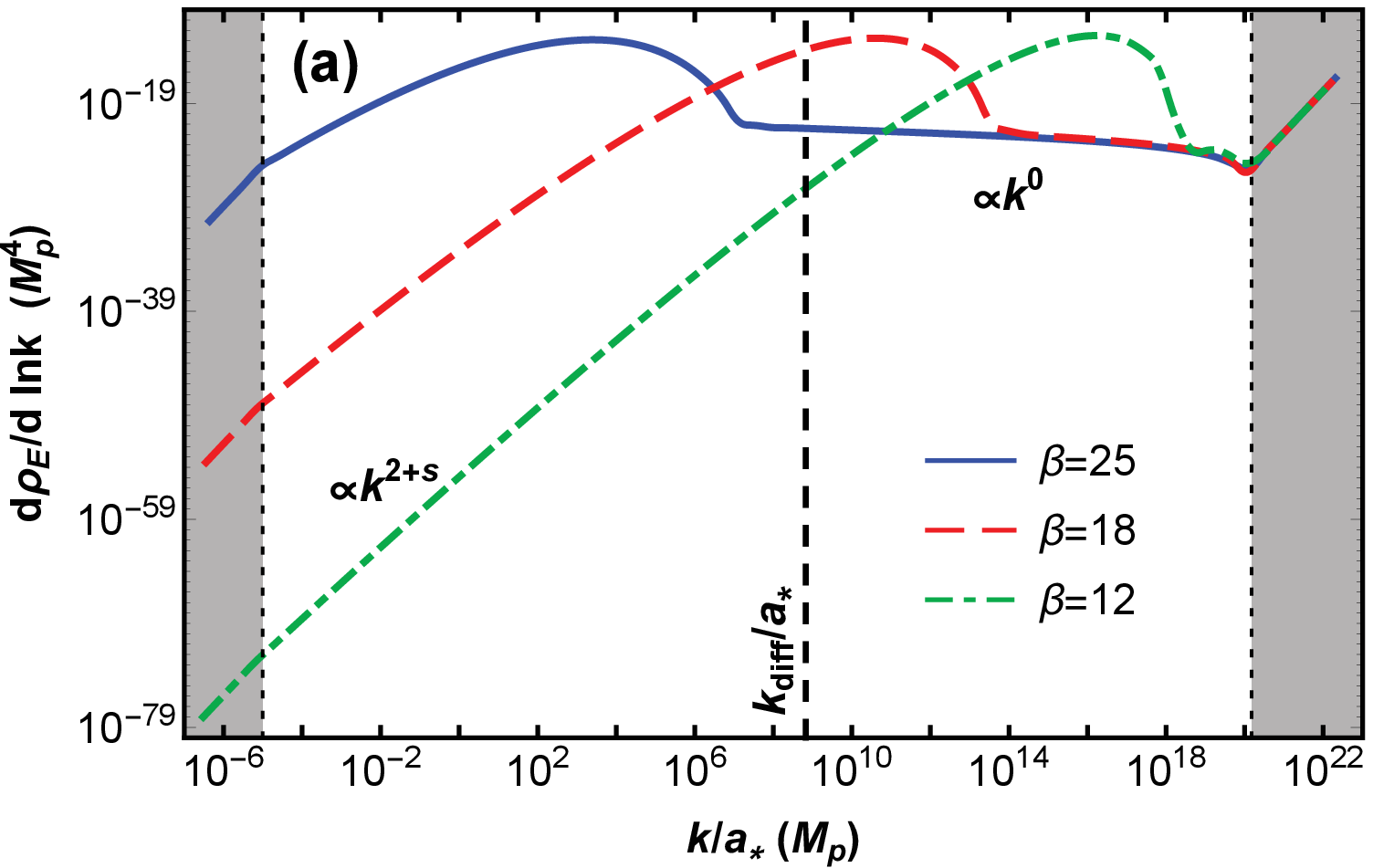}
	\includegraphics[width=0.49\textwidth]{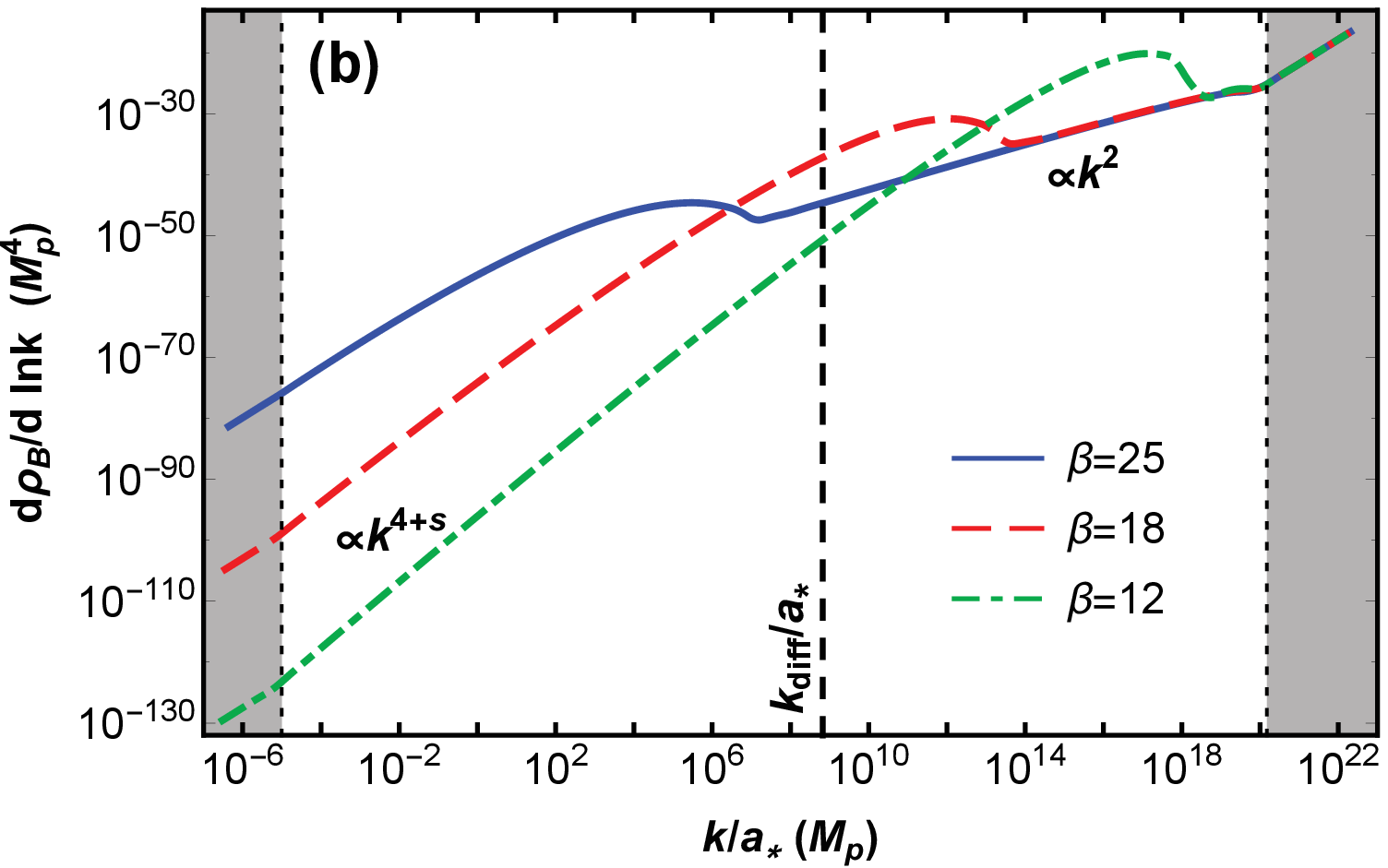}
	\caption{The electric (a) and magnetic (b) power spectra in the model with Ratra coupling function $f=\exp(\beta\phi/M_{p})$ for $\beta=12,\ 18$, and $25$. Left shaded regions correspond to the modes which are outside the horizon even before the beginning of inflation. Right shaded regions correspond to the modes which have not crossed the horizon even until the end of inflation and, hence, do not undergo enhancement. The vertical black dashed lines show the momentum which corresponds to the cosmic diffusion scale $k_{\rm diff}/a_{\ast}$.
		\label{fig-spectrum-BR-Ratra}}
\end{figure}

Using the solutions of Eqs.~(\ref{Friedmann}), (\ref{KGF-2}), and (\ref{Maxwell-4}) as a background we numerically solve Eq.~(\ref{mode-eq}) for the electromagnetic mode function for all modes which cross the horizon during inflation and calculate the power spectra of generated electromagnetic fields. They are shown in Fig.~\ref{fig-spectrum-BR-Ratra} for three different values of the parameter $\beta$. As it was mentioned above, in the early stages of inflation the electric energy density is small and it does not cause the backreaction. For all the modes which cross the horizon at this stage the spectra have the same scaling with $k$ as unperturbed power spectra described by Eqs.~(\ref{E-spectrum-Ratra}) and (\ref{B-spectrum-Ratra}); however, their amplitudes are damped in comparison to the free case due to the backreaction which occurs later. When the coupling function starts to change faster, the power spectra undergo significant enhancement. This is reflected in the peaks in power spectra in Fig.~\ref{fig-spectrum-BR-Ratra} and this also corresponds to the fast increase of the electric and magnetic integral energy densities, which is shown in Fig.~\ref{fig-magnetic-Ratra}. Finally, when the backreaction regime occurs the power spectra behave as predicted in Eq.~(\ref{spectra-br-regime}). For larger values of $\beta$ the backreaction occurs earlier and the range of modes, which behave in such a manner, is wider.

\begin{figure}[h!]
	\centering
	\includegraphics[width=0.49\textwidth]{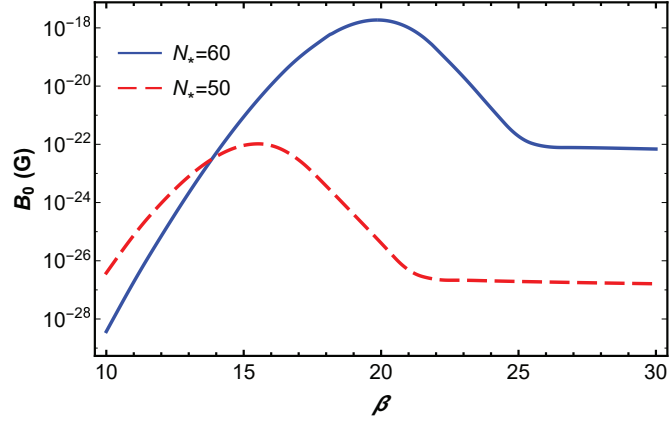}
	\caption{The dependence of the present-day value of the magnetic field on parameter $\beta$ for two numbers of $e$-folds from the moment of the pivot scale horizon crossing to the end of inflation, $N_{\ast}=50$ (red dashed curve) and $N_{\ast}=60$ (blue solid curve).
		\label{fig-magnetic-today-Ratra}}
\end{figure}

Finally, from the known power spectra we can calculate the present-day value of the magnetic field using Eq.~(\ref{magnetic-field-today}). Figure~\ref{fig-magnetic-today-Ratra} shows its dependence on $\beta$ for two values of the number of $e$-folds from the pivot scale horizon crossing to the end of inflation, $N_{\ast}=50$ (red dashed line) and $N_{\ast}=60$ (blue solid line). Note that not all modes make their contributions to the present value of the magnetic field since not all of them can survive the cosmic diffusion at the later stages of the evolution of the Universe.

For low values of the parameter $\beta\lesssim 10$ the backreaction does not occur at all and the power spectra behave like in the unperturbed case considered in Ref.~\cite{Vilchinskii:2017}. The magnetic field shows a steep growing dependence on $\beta$. However, for larger values of $\beta$ the backreaction has an impact on the spectra and the growth slows down. The magnetic field has maximal value in the case when the peak in the magnetic power spectrum occurs for modes in the vicinity of the cosmic diffusion scale $k_{\rm diff}$. For $N_{\ast}=60$ this happens for $\beta=18-22$. For larger values of $\beta$ the backreaction regime sets much before the diffusion scale crosses the horizon and the modes which contribute to the magnetic field have power spectrum $\propto k^{2}$. Therefore, the dependence of $B_{0}$ tends to a constant given by Eq.~(\ref{B0}). This can be seen as plateaus for large $\beta$ in Fig.~\ref{fig-magnetic-today-Ratra}.

Figure~\ref{fig-magnetic-today-Ratra} shows that the generated magnetic field can be as large as $10^{-18}$~G. However, its blue spectrum with $n_{B}=4+s$ for long wavelength modes and $n_{B}\approx 2$ for the shortest ones implies that the coherence length is very small and is comparable with the cosmic diffusion scale, i.e. 1~A.U. For such magnetic fields the lower bound required by the observations \cite{Neronov:2010,Taylor:2011,Dermer:2011,Caprini:2015} is $B_{\rm min}\sim 10^{-18} (1\,{\rm A.U.}/1\,{\rm Mpc})^{-1/2}\,{\rm G}\sim 5\cdot 10^{-13}\,{\rm G}$. Thus, the theoretical prediction gives much smaller value of the magnetic field than that experimentally observed.

\section{Conclusions}
\label{sec-concl}

In this work, we have studied how the backreaction of electric fields and the Schwinger effect influence the inflationary magnetogenesis in the model with the standard kinetic coupling of the electromagnetic field to the inflaton $f^{2}(\phi)F_{\mu\nu}F^{\mu\nu}$ and decreasing coupling function $f(\phi)$. Such a model leads to the generation of strong electric fields which dominate over the magnetic ones and backreact on the inflationary dynamics. We derived the self-consistent system of equations which describes the joint evolution of the scale factor, inflaton field, and electric field during inflation. We took into account also the cosmological Schwinger effect which turned out to be important at the end of inflation, helping to finish the inflation stage and providing the mechanism of reheating which is complementary to the usual scenario with fast inflaton oscillations. To the best of our knowledge, such a self-consistent description has not been studied in the literature before.

In our analysis of the electromagnetic field evolution, it was assumed that the modes inside the horizon are described by the quantum electromagnetic field state in the Bunch-Davies vacuum. The modes far outside the horizon can be described classically, contributing to the electric and magnetic energy densities which can be associated with the observed large-scale fields. As the Universe expands, more and more modes cross the horizon and contribute. We took this fact into account through an additional boundary term in the equation for the electric energy density. The role of this term is rather subtle. Since the equation for the superhorizon modes is classical and homogeneous, it would imply only the trivial solution in the absence of this term which describes the quantum-to-classical transition. 

The backreaction becomes important when the energy density of the electric field becomes as large as the product $\epsilon \rho_{\rm inf}$ of the slow-roll parameter $\epsilon$ and the inflaton energy density $\rho_{\rm inf}$. After this, the electric energy density remains almost constant until the end of inflation. The presence of such backreacting electric fields slows down the rolling of the inflaton and prolongs inflation. For the modes that cross the horizon after the backreaction becomes important, the electric power spectrum is scale-invariant and the magnetic one has a blue tilt with the spectral index $n_{B}=2$. This behavior does not depend on the explicit form of the coupling function.

Although the Schwinger effect naively should be very important in the presence of electric fields, we found that this is not the case. The point is that the Schwinger effect is tuned by the value of the effective charge $e_{\rm eff}=e/f$. At the beginning of inflation when the electric field is weak and the coupling function is very large, $f\gg 1$, the Schwinger effect is not efficient and has no impact on the evolution of the inflaton and the electromagnetic fields. It becomes important at the late stage when the inflaton field is close to the potential minimum and the coupling function $f\to 1$. In this case, the Schwinger current leads to a quick dissipation of the electric field due to the high conductivity of the produced plasma. The magnetic field becomes ``frozen-in'' and evolves in accordance with the flux conservation law. The energy density of the produced particles becomes comparable with the inflaton energy density. Therefore, the Schwinger effect starts to reheat the Universe even before the fast oscillations of the inflaton. Thus, we confirm that the Schwinger reheating previously discussed in Ref.~\cite{Tangarife:2017} can help in filling the Universe with particles.

However, this does not necessarily mean the completion of reheating and does not exclude the possibility of the standard reheating mechanisms, such as parametric resonance processes (narrow, broad, and stochastic) \cite{Kofman:1997} or perturbative decay of the inflaton, because even after the Schwinger effect turned on the inflaton field still oscillates in its potential minimum. To find out what mechanism is the most effective one has to extend the time evolution to the preheating stage and to include the interaction with other fields.
However, the applicability of our theory is not well justified during the preheating stage. First of all, since the electric energy density has been strongly reduced due to the Schwinger process, the large field approximation for the Schwinger current given by Eq.~(\ref{strong-field}) is no longer correct. The low field expressions (\ref{weak-field-bose})--(\ref{weak-field-fermi}) depend on the renormalization scheme and have, according to the studies in the literature, such features as the negative conductivity or the infrared hyperconductivity which are rather questionable from the physical point of view. Finally, expressions (\ref{weak-field-bose})-(\ref{weak-field-fermi}) were obtained in the de Sitter space-time for the constant electric field and cannot be used during the preheating stage. Therefore, an appropriate study of the Schwinger reheating requires the use of the correct expressions for the Schwinger current.

In our analysis, we used the expressions for the Schwinger current obtained in the case of a constant and homogeneous electric field in de Sitter space-time \cite{Kobayashi:2014,Hayashinaka:2016a}. In Ref.~\cite{Kitamoto:2018} it was shown that in the quasiclassical approximation the current has a behavior similar to that in the case of a constant electric field, which implies that our analysis is correct at least qualitatively. Nevertheless, the most accurate approach describing the self-consistent evolution of the electric field and particles produced due to the Schwinger effect is possible to realize only in the framework of the kinetic theory. However, this lies far away from the scope of our article and should be addressed elsewhere.

The magnetic field energy density is subdominant compared to the electric one. Therefore, one can neglect it when solving the background equations. Using the inflaton and scale factor time dependencies, we integrate the equation for the electromagnetic field mode function and compute the magnetic power spectrum. Such a perturbative approach can be justified because the calculated magnetic energy density is much less than the background electric energy density. Having determined the magnetic power spectrum we estimated the value of the magnetic field at the present epoch. We took into account that after the end of inflation the magnetic field evolves according to the flux conservation law, i.e. $B\propto a^{-2}$, and integrated the power spectrum over all modes, which can survive the cosmic diffusion during the late stages on Universe evolution. The last mode which gives the contribution has the wavelength $\sim 1\,{\rm A.U.}$ at the present time \cite{Grasso:2001}.

Were the Schwinger effect absent, the evolution of the magnetic field energy density could be determined perturbatively during the preheating stage as it was done during inflation. Later, during the reheating stage the Universe becomes filled with all sorts of particles and particularly charged ones. Therefore, due to high conductivity the electric field disappears and the magnetic one starts to evolve like  $B\propto a^{-2}$. The role of the Schwinger effect, therefore, is to turn on this regime earlier, near the end of inflation. As a result, the present-day value of the magnetic field may be reduced at most by a factor of $e^{-2\Delta N_{r}}$, where $\Delta N_{r}$ is the number of $e$-folds between the end of inflation and the beginning of reheating, i.e., the duration of preheating.

We studied numerically the generation of electromagnetic fields in the Starobinsky inflationary model for two types of coupling functions: (i) the powerlike $f\propto a^{\alpha}$ and (ii) the Ratra-type $f=\exp(\beta\phi/M_{p})$. 
Magnetogenesis in the kinetic coupling model with a powerlike coupling function was considered previously in Ref.~\cite{Demozzi:2009}, where it was shown that the backreaction takes place for $\alpha<-2.2$ and should be included self-consistently in analysis. We showed that the backreaction becomes important when $\rho_{E}\sim \epsilon\rho_{\rm inf}\ll \rho_{\rm inf}$ and the unperturbed calculation becomes inapplicable even earlier, for $\alpha<-2.1$. Moreover, we extended the analysis for all negative values of $\alpha$ and showed that the present-day value of the magnetic field cannot exceed $10^{-22}\,$G, which is still too small to explain the observational data. We numerically confirm the results of Ref.~\cite{Kanno:2009} showing that for all $\alpha<-2$ in the backreaction regime the electromagnetic power spectra are similar to the case $\alpha=-2$ (the so-called attractor) and this is the reason for strong suppression of the magnetic field.

For the second type of coupling function, the rapid enhancement of the electromagnetic fields occurs only at the late stage of inflation when the inflaton field starts to deviate from the slow-roll regime. Although the spectral index for the modes which contribute to the observed value of the magnetic field is determined by the free dynamics without backreaction and is equal to $n_{B}=6-\sqrt{6}\beta/N_{e}$, the amplitude of the spectrum is damped due to the backreaction. In the most favorable situation, when the physically relevant modes cross the horizon at the time when the coupling function decreases faster than $a^{-2}$, the strength of the present-day magnetic field can reach $\sim 10^{-18}\,$G. However, the spectrum of this field has a blue tilt that leads to a small coherence length comparable with the cosmic diffusion scale, i.e. $\sim 1\,{\rm A.U.}$ For such magnetic fields the lower bound required by the observations \cite{Neronov:2010,Taylor:2011,Dermer:2011,Caprini:2015} is $B_{\rm min}\sim 10^{-18} (1\,{\rm A.U.}/1\,{\rm Mpc})^{-1/2}\,{\rm G}\sim 5\cdot 10^{-13}\,{\rm G}$. Therefore, the theoretical prediction lies well below the observational limit.

It is known that the electric fields produced in magnetogenesis models induce the curvature perturbations during inflation and they can give tight constraints on the models from the CMB observations. These constraints on the kinetic coupling models have been studied in the literature (e.g. \cite{Barnaby:2012,Fujita:2013}). Since the electric spectrum has a large power on the CMB scale in Fig.~3, this case may be excluded by this argument. However, the calculation of the curvature perturbation power spectrum in the strong backreaction regime deserves a separate investigation and lies outside the scope of our article.

In our work, we considered only the case of kinetic coupling between the inflaton and the electromagnetic field and showed that the backreaction from the electric fields strongly suppresses magnetogenesis. This makes the kinetic coupling model unfavorable in contrast to the case of axion coupling of the pseudoscalar inflaton $\varphi$ via the term $\mathcal{L}\propto \varphi F_{\mu\nu}\tilde{F}^{\mu\nu}$, where $\tilde{F}^{\mu\nu}$ is the dual electromagnetic tensor \cite{Durrer:2011}. This case seems to be promising because it generates the helical magnetic fields \cite{Garretson:1992,Durrer:2011,Anber:2006,Caprini:2014}, which are more stable against dissipation and their coherence length can be increased due to the inverse cascade process \cite{Joyce:1997,Cornwall:1997,Boyarsky:2012,Sydorenko:2016,Boyarsky:2015,Gorbar:2016a,Gorbar:2016b}. Nevertheless, this scenario also suffers from the backreaction problem \cite{Domcke:2018} and it would be interesting to implement our equations in this case. We plan to address this issue in future studies.

\vspace{5mm}

\begin{acknowledgments}
	The work of S.~I.~V. is supported partially by the Swiss National Science Foundation, Grant No. SCOPE IZ7370-152581. 
	S.~I.~V. is grateful to the Swiss National Science Foundation (individual Grant No. IZKOZ2-154984). The work of S.~I.~V. and O.~O.~S. is supported by the Department of Targeted Training of Taras Shevchenko National University of Kyiv under the National Academy of Sciences of Ukraine, project 6F-2017. 
	S.~I.~V. is grateful to Professor Marc Vanderhaeghen and Dr.\@ Vladimir Pascalutsa for their support and kind hospitality at the Institut f\"{u}r Kernphysik,
	Johannes Gutenberg-Universit\"{a}t Mainz, Germany, where part of this work was done.
	The work of E.V.G. was supported partially by the Ukrainian State Foundation for Fundamental Research.
	The work of O.~O.~S. was supported in part by the European Research Council, Grant No. AdG-2015 694896.
	O.~O.~S. is grateful to Professor Mikhail Shaposhnikov for his kind hospitality at the Institute of Physics, \'{E}cole Polytechnique F\'{e}d\'{e}rale de Lausanne, Switzerland, where the final part of this work was done.
\end{acknowledgments}

\end{document}